\documentclass[letterpaper,twocolumn,10pt]{article}
\usepackage{usenix,epsfig,endnotes}
\usepackage{graphicx}
\usepackage{balance}
\usepackage{times}
\usepackage{epsfig,endnotes}
\usepackage{epstopdf}
\usepackage[font={footnotesize,it}]{caption}
\usepackage{subcaption}
\usepackage[rgb]{xcolor} 
\usepackage{url}
\usepackage{amsmath}
\usepackage[normalem]{ulem} 
\usepackage{fancyvrb} 
\usepackage{float, lipsum}
\usepackage{listings}
\usepackage{soul}
\usepackage[T1]{fontenc}
\usepackage{mwe}    
\usepackage[utf8]{inputenc}
\usepackage{cleveref}
\usepackage{pbox}
\usepackage{mathptmx}
\usepackage[11pt]{moresize}
\usepackage{xcolor,colortbl}
\usepackage{array}
\usepackage{booktabs}
\usepackage{enumitem}
\setlist{leftmargin=4.0mm}

\usepackage{mathtools}
\usepackage{titlesec}
\DeclarePairedDelimiter{\ceil}{\lceil}{\rceil}

\definecolor{tableHeaderColor}{rgb}{0.57, 0.64, 0.81}
\definecolor{lightRowColor}{rgb}{1,1,1}
\definecolor{darkRowColor}{rgb}{0.8,0.8,0.8}
\definecolor{cellLightHighlightColor}{rgb}{0.74, 0.85, 0.34}
\definecolor{cellDarkHighlightColor}{rgb}{0.74, 0.85, 0.34}
\definecolor{cellHighlightTotalColor}{rgb}{0.94, 0.95, 0.64}

\crefname{section}{section}{section}
\Crefname{section}{Section}{Section}

\DeclareCaptionFormat{myformat}{#1#2#3\vspace{-2.5mm}\hrulefill\vspace{-2mm}}
\titlespacing*{\section}
{0pt}{1.0ex}{0ex}
\titlespacing*{\subsection}
{0pt}{1.0ex}{0ex}
\titlespacing*{\subsubsection}
{0pt}{1.0ex}{0ex}

\begin{document}

\title{HopsFS: Scaling Hierarchical File System Metadata Using NewSQL Databases}

\author{
{\rm Salman Niazi, Mahmoud Ismail,}\\{\rm Seif Haridi, Jim Dowling}\\
KTH - Royal Institute of Technology \\
{\small\{smkniazi, maism, haridi, jdowling\}@kth.se}
\and
{\rm Steffen Grohsschmiedt}\\
Spotify AB \\
{\small steffeng@spotify.com} \\ \\
\and
{\rm Mikael Ronstr{\"o}m}\\
Oracle \\
{\small mikael.ronstrom@oracle.com} 
} 
%
\maketitle

\hyphenation{data-node}
\hyphenation{data-nodes}
\hyphenation{data-base}
\hyphenation{i-node}
\hyphenation{i-nodes}
\hyphenation{Name-Node}
\hyphenation{Name-Nodes}
\hyphenation{name-node}
\hyphenation{name-nodes}
\hyphenation{acc-ess-ed}

\subsection*{Abstract}

Recent improvements in both the performance and scalability of shared-nothing, transactional, in-memory NewSQL databases have reopened the research question of whether distributed metadata for hierarchical file systems can be managed using commodity databases. In this paper, we introduce HopsFS, a next generation distribution of the Hadoop Distributed File System (HDFS) that replaces HDFS' single node in-memory metadata service, with a distributed metadata service built on a NewSQL database. By removing the metadata bottleneck, HopsFS enables an order of magnitude larger and higher throughput clusters compared to HDFS. Metadata capacity has been increased to at least 37 times HDFS' capacity, and in experiments based on a workload trace from Spotify, we show that HopsFS supports 16 to 37 times the throughput of Apache HDFS. HopsFS also has lower latency for many concurrent clients, and no downtime during failover. Finally, as metadata is now stored in a commodity database, it can be safely extended and easily exported to external systems for online analysis and free-text search.









\section{Introduction}

Distributed file systems are an important infrastructure component of many large scale data-parallel processing systems, such as MapReduce~\cite{DeanMR}, Dryad~\cite{isardDryad}, Flink~\cite{AlexandrovStratosphere} and Spark~\cite{ZahariaSpark}. By the end of this decade, data centers storing multiple exabytes of data will not be uncommon~\cite{EMC2_Hadoop, Exabyte_Cern}. For large distributed hierarchical file systems, the metadata management service is the scalability bottleneck~\cite{shvachkoHdfsLimitations}. Many existing distributed file systems store their metadata on either a single node or a shared-disk file systems, such as storage-area network (SAN), both of which have limited scalability. Well known examples include GFS~\cite{GFS-2003}, HDFS~\cite{shvachko_hdfs}, QFS~\cite{qfs}, Farsite~\cite{farsite1}, Ursa Minor~\cite{Ursa_Minor}, GPFS~\cite{GPFS}, Frangipani~\cite{Frangipani}, GlobalFS~\cite{GlobalFS}, and Panasas~\cite{Panasas}. Other systems scale out their metadata by statically sharding the namespace and storing the shards on different hosts, such as NFS~\cite{NFS}, AFS~\cite{AFS}, MapR~\cite{mapr},  Locus~\cite{LOCUS-GJ-Popek}, Coda~\cite{CODA}, Sprite~\cite{Sprite} and XtreemFS~\cite{xtreemfs}. However, statically sharding the namespace negatively affects file system operations that cross different shards, in particular \emph{move} operation. Also, it complicates the management of the file system, as administrators have to map metadata servers to namespace shards that change in size over time.

Recent improvements in both the performance and scalability of shared-nothing, transactional, in-memory NewSQL~\cite{newsql14sigmod} databases have reopened the possibility of storing  distributed file system metadata in a commodity database. To date, the conventional wisdom has been that it is too expensive (in terms of throughput and latency) to store hierarchical file system metadata fully normalized in a distributed database~\cite{Seltzer2009,LevyDfsServey}. 


In this paper we show how to build a high throughput and low operational latency distributed file system using a NewSQL database. We present HopsFS, a new distribution of the Hadoop Distributed File System (HDFS)~\cite{shvachko_hdfs}, which decouples file system metadata storage and management services. HopsFS stores all metadata normalized in a highly available, in-memory, distributed, relational database called Network Database (NDB), a NewSQL storage engine for MySQL Cluster~\cite{mySQLCluster,Ronstroem2005}. HopsFS provides redundant stateless servers (namenodes) that in parallel, read and update metadata stored in the database.


HopsFS encapsulates file system operations in distributed transactions. To improve the performance of file system operations, we leverage both classical database techniques such as \textit{batching} (bulk operations) and \textit{write-ahead} caches within transactions, as well as distribution aware techniques commonly found in NewSQL databases. These distribution aware NewSQL techniques include \textit{application defined partitioning} (we partition the namespace such that the metadata for all immediate descendants of a directory (child files/directories) reside on the same database shard for efficient directory listing),  and \textit{distribution aware transactions} (we start a transaction on the database shard that stores all/most of the metadata required for the file system operation), and \textit{partition pruned index scans} (scan operations are localized to a single database shard~\cite{partitionPruning}). We also introduce an inode hints cache for faster resolution of file paths. Cache hits when resolving a path of depth \emph{N} can reduce the number of database round trips from \emph{N}~to~\emph{1}.

However, some file system operations on large directory subtrees (such as \emph{move}, and \emph{delete}) may be too large to fit in a single database transaction. For example, deleting a folder containing millions of files cannot be performed using a single database transaction due to the limitations imposed by the database management system on the number of operations that can be included in a single transaction. For these \textit{subtree operations}, we introduce a novel protocol that uses an application level distributed locking mechanism to isolate large subtrees to perform file system operations. After isolating the subtrees large file system operations are broken down into smaller transactions that execute in parallel for performance. The subtree operations protocol ensures that the consistency of the namespace is not violated if the namenode executing the operation fails.

HopsFS is a drop-in replacement for HDFS. HopsFS has been running in production since April 2016, providing Hadoop-as-a-Service for researchers at a data center in Lule\aa{}, Sweden~\cite{hopsLulea}. In experiments, using a real-world workload generated by Hadoop/Spark applications from Spotify, we show that HopsFS delivers 16 times higher throughput than HDFS, and HopsFS has no downtime during failover. For a more write-intensive workload, HopsFS delivers 37 times the throughput of HDFS. To the best of our knowledge HopsFS is the first open-source distributed file system that stores fully normalized metadata in a distributed relational database.

\section{Background}
\label{sec:hopsfs-architecture}
This section describes Hadoop Distributed File System (HDFS) and MySQL Cluster Network Database (NDB) storage engine. 
\subsection{Hadoop Distributed File System}
\begin{figure}[!tb]
        \centering
        \includegraphics[width=0.48\textwidth]{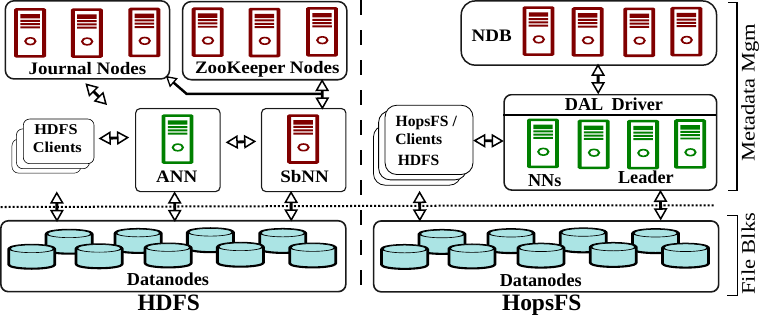}
        \caption{System architecture for HDFS and HopsFS. For high availability, HDFS requires an Active NameNode (ANN), at least one Standby NameNode (SbNN), at least three Journal Nodes for quorum-based replication of the write ahead log of metadata changes, and at least three ZooKeeper instances for quorum based coordination. HopsFS supports multiple stateless namenodes that access the metadata stored in NDB database nodes.}    
\label{fig:hopsfs-arch}
\end{figure}

The Hadoop Distributed File System (HDFS)~\cite{shvachko_hdfs} is an open source implementation of the Google File System~\cite{GFS-2003}. HDFS' metadata is stored on the heap of single Java process called the Active NameNode (ANN), see Figure~\ref{fig:hopsfs-arch}. The files are split into small (typically 128 MB) blocks that are by default triple replicated across the datanodes. For high availability of the metadata management service, the Active namenode logs changes to the metadata to journal servers using quorum based replication. The metadata change log is replicated asynchronously to a Standby NameNode (SbNN), which also performs checkpointing functionality. In HDFS, the ZooKeeper coordination service~\cite{zookeeper} enables both agreement on which machine is running the active namenode (preventing a split-brain scenario) as well as coordinating failover from the active to the standby namenode.

The namenode serves requests from potentially thousands of datanodes and clients, and keeps the metadata strongly consistent by executing the file system operations atomically. The namenode implements atomic operations using a single global lock on the entire file system metadata, providing single-writer, multiple-readers concurrency semantics. Some large file system operations are not atomic, as they would hold the global lock for too long, starving clients. For example, deleting large directories is performed in batches, with inodes first being deleted, then the blocks are deleted in later phases. Moreover, as writing namespace changes to the quorum of journal nodes can take long time, the global file system lock is released before the operation is logged to prevent other clients from starving. Concurrent clients can acquire the file system lock before the previous operations are logged, preventing starvation, at the cost of inconsistent file system operations during namenode failover. For example, when the active namenode fails all the changes that are not logged to the journal nodes will be lost.

The datanodes are connected to both active and standby namenodes. All the datanodes periodically generate a block report containing information about its own stored blocks. The namenode processes the block report to validate the consistency of the namenode's blocks map with the blocks actually stored at the datanode.

In HDFS the amount of metadata is quite low relative to file data. There is approximately 1 gigabyte of metadata for every petabyte of file system data~\cite{shvachkoHdfsLimitations}. Spotify's HDFS cluster has $1600{\raisebox{1.8pt}{\scriptsize$+$}}$ nodes, storing 60 petabytes of data, but its metadata fits in 140 gigabytes Java Virtual Machine (JVM) heap. The extra heap space is taken by temporary objects, RPC request queues and secondary metadata required for the maintenance of the file system. However, current trends are towards even larger HDFS clusters (Facebook has HDFS clusters with more than 100 petabytes of data~\cite{polato2014comprehensive}), but current JVM garbage collection technology does not permit very large heap sizes,  as the application pauses caused by the JVM garbage collector affects the operations of HDFS~\cite{hdfsGCPausesJira}. As such, JVM garbage collection technology and the monolithic architecture of the HDFS namenode are now the scalability bottlenecks for Hadoop~\cite{shvachkoHdfsLimitations}. Another limitation with this architecture is that data structures are optimized to reduce their memory footprint with the result that metadata is difficult to modify or export to external systems.

\subsection{Network Database (NDB)}
MySQL Cluster is a shared-nothing, replicated, in-memory, auto-sharding, \emph{consistent}, NewSQL relational database~\cite{mySQLCluster}. Network DataBase (NDB) is the storage engine for MySQL Cluster. NDB supports both datanode-level and cluster-level failure recovery. The datanode-level failure recovery is performed using transaction redo and undo logs. NDB datanodes also asynchronously snapshot their state to disk to bound the size of logs and to improve datanode recovery time. Cluster-level recovery is supported using a global checkpointing protocol that increments a global epoch-ID, by default every 2 seconds. On cluster-level recovery, datanodes recover all transactions to the latest epoch-ID. 

NDB horizontally partitions the tables among storage nodes called NDB datanodes. NDB also supports application defined partitioning (ADP) for the tables. Transaction coordinators are located at all NDB datanodes, enabling high performance transactions between data shards, that is, multi-partition transactions. Distribution aware transactions (DAT) are possible by providing a \textit{hint}, based on the application defined partitioning scheme, to start a transaction on the NDB datanode containing the data read/updated by the transaction. In particular, single row read operations and partition pruned index scans (scan operations in which a single data shard participates) benefit from distribution aware transactions as they can read all their data locally~\cite{partitionPruning}. Incorrect hints result in additional network traffic being incurred but otherwise correct system operation.

\subsubsection{NDB Data Replication and Failure Handling}
\label{ndb-data-replication}
NDB datanodes are organized into node groups, where the data replication factor, $R$, determines the number of datanodes in a node group. Given a cluster size $N$, there are $N/R$ node groups.  NDB partitions tables (hash partitioning by default) into a fixed set of partitions distributed across the node groups. New node groups can be added online, and existing data is automatically rebalanced to the new node group. A partition is a fragment of data stored and replicated by a node group. Each datanode stores a copy (replica) of the partition assigned to its node group. In NDB, the default replication degree is two, which means that each node group can tolerate one NDB datanode failure as the other NDB datanode in the node group contains a full copy of the data. So, a twelve node NDB cluster has six node groups can tolerate six NDB datanode failures as long as there is one surviving NDB datanode in each of the node groups. To tolerate multiple failures within a node group, the replication degree can be increased at the cost of lower throughput.   

\subsubsection{Transaction Isolation}
NDB only supports \emph{read-committed} transaction isolation, which guarantees that any data read is committed at the moment it is read. The \emph{read-committed} isolation level does not allow \emph{dirty} reads but \emph{phantom} and \emph{fuzzy} (non-repeatable) reads can happen in a transaction~\cite{txIosLvlCritique}. However, NDB supports row level locks, such as, \emph{exclusive} (write) locks, \emph{shared} (read) locks, and \emph{read-committed} locks that can be used to isolate conflicting transactions.

\section{HopsFS Overview}

HopsFS is a fork of HDFS  v2.0.4.  Unlike HDFS, HopsFS provides a scale-out metadata layer by decoupling the metadata storage and manipulation services. HopsFS supports multiple stateless namenodes, written in Java, to handle clients' requests and process the metadata stored in an external distributed database, see Figure~\ref{fig:hopsfs-arch}. Each namenode has a Data Access Layer (DAL) driver that, similar to JDBC, encapsulates all database operations allowing HopsFS to store the metadata in a variety of NewSQL databases. The internal management (housekeeping) operations, such as datanode failure handling, must be coordinated amongst the namenodes. HopsFS solves this problem by electing a leader namenode that is responsible for the housekeeping. HopsFS uses the database as shared memory to implement a leader election and membership management service. The leader election protocol assigns a unique ID to each namenode, and the ID of the namenode changes when the namenode restarts. The leader election protocol defines an alive namenode as one that can write to the database in bounded time, details for which can be found in~\cite{salmanLE}.

Clients can choose between \emph{random}, \emph{round-robin}, and \emph{sticky} policies for selecting a namenode on which to execute file system operations. HopsFS clients periodically refresh the namenode list, enabling new namenodes to join an operational cluster. 
HDFS v2.x clients are fully compatible with HopsFS, although they do not distribute operations over namenodes, as they assume there is a single active namenode. Like HDFS, the datanodes are connected to all the namenodes, however, the datanodes send the block reports to only one namenode. The leader namenode load balances block reports over all alive namenodes.

In~\cref{sec:metadata-hierarchical}, we discuss how HopsFS' auto sharding scheme enables common file system operations to read metadata using low cost database access queries.~\Cref{sec-detailed-solution} discusses how the consistency of the file system metadata is maintained by converting file system operations into distributed transactions, and how the latency of the distributed transactions is reduced using per-transaction and namenode level caches. Then, in ~\cref{sec-subtree-ops}, a protocol is introduced to handle file system operations that are too large to fit in a single database transaction.

\begin{table}[tb]
\scriptsize
\centering
\begin{tabular}{|p{0.105\textwidth} p{0.09\textwidth}|p{0.11\textwidth} p{0.098\textwidth}|}
\hline
\hline
\cellcolor{tableHeaderColor}\textbf{Op Name} & \cellcolor{tableHeaderColor}\textbf{Percentage}  & \cellcolor{tableHeaderColor}\textbf{Op Name}  & \cellcolor{tableHeaderColor}\textbf{Percentage }\\ 
\hline
\hline

\rowcolor{lightRowColor}append file                         &  0.0\%                                                                                   &  content  summary                   &   0.01\%   \\
\rowcolor{darkRowColor}mkdirs                                &  0.02\%                                                                                 &  set permissions                       &   0.03\% [26.3\%\raisebox{1.8pt}{\scriptsize$*$}] \\
\rowcolor{lightRowColor}set replication                    &  0.14\%                                                                                 &   set owner                                &    0.32 \%  [100\%\raisebox{1.8pt}{\scriptsize$*$}] \\
\rowcolor{darkRowColor}delete	                             &  0.75\% [3.5\%\raisebox{1.8pt}{\scriptsize$*$}]                &  create file	                            &    1.2\% \\
\rowcolor{lightRowColor}move 	                             &  1.3\%  [0.03\%\raisebox{1.8pt}{\scriptsize$*$}]               &  add blocks                                &   1.5\%  \\
\rowcolor{darkRowColor}\cellcolor{cellDarkHighlightColor}\textbf{list (\emph{listStatus})}          &  \cellcolor{cellDarkHighlightColor}\textbf{9\% [94.5\%\raisebox{1.8pt}{\scriptsize$*$}]}	    &  \cellcolor{cellDarkHighlightColor}\textbf{stat (\emph{fileInfo})}          &   \cellcolor{cellDarkHighlightColor}\textbf{17\% [23.3\%\raisebox{1.8pt}{\scriptsize$*$}]}	\\		
\rowcolor{lightRowColor}\cellcolor{cellDarkHighlightColor}\textbf{read (\emph{getBlkLoc})}             & \cellcolor{cellDarkHighlightColor} \textbf{68.73\%}		 &\cellcolor{cellHighlightTotalColor}\textbf{Total Read Ops}      & \cellcolor{cellHighlightTotalColor}\textbf{94.74\%} \\

\hline
\hline
\end{tabular}
\caption{Relative frequency of file system operations for Spotify's HDFS cluster. \textit{List}, \textit{read}, and \textit{stat} operations account for $\approx 95\%$ of the metadata operations in the cluster. \\ \raisebox{1.8pt}{\scriptsize$*$}Of which, the relative percentage is on directories}
\label{Operation-Percentages}
\vspace{2mm}
\hrulefill
\end{table}


\section{HopsFS Distributed Metadata}
\label{sec:metadata-hierarchical}
Metadata for hierarchical distributed file systems typically contains information on inodes, blocks, replicas, quotas, leases and mappings (directories to files, files to blocks, and blocks to replicas). When metadata is distributed, an application defined partitioning scheme is needed to shard the metadata and a consensus protocol is required to ensure metadata integrity for operations that cross shards. Quorum-based consensus protocols, such as Paxos, provide high performance within a single shard, but are typically combined with transactions, implemented using the two-phase commit protocol, for operations that cross shards, as in Megastore~\cite{megastore11} and Spanner~\cite{spanner}. File system operations in HopsFS are implemented primarily using multi-partition transactions and row-level locks in MySQL Cluster to provide serializability~\cite{Herlihy90} for metadata operations.

\begin{figure}[t]
 \centering
 \includegraphics[width=0.45\textwidth]{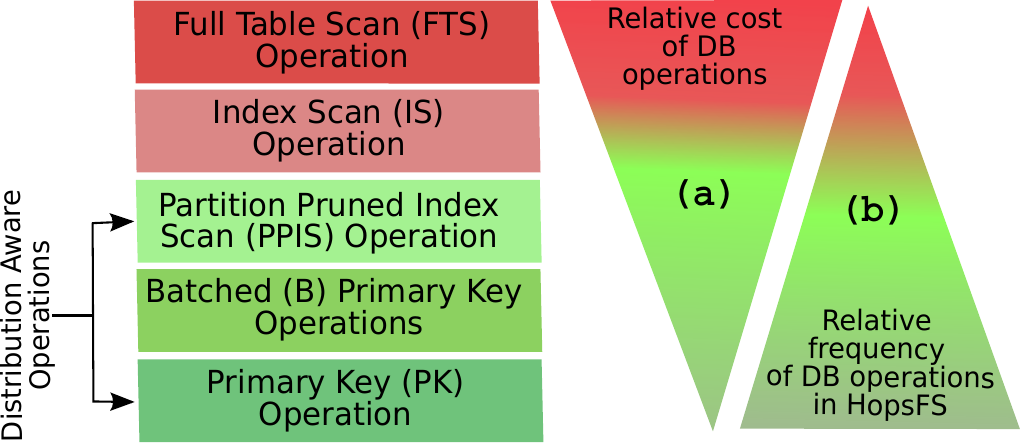}
 \caption{(a) Shows the relative cost of different operations in NewSQL database. (b) HopsFS avoids FTS and IS operations as the cost these operation is relatively higher than PPIS, B, and PK operations. }
 \label{fig:pyramid}
\end{figure}

The choice of partitioning scheme for the hierarchical namespace is a key design decision for distributed metadata architectures. We base our partitioning scheme on the expected relative frequency of HDFS operations in production deployments and the cost of different database operations that can be used to implement the file system operations. Table~\ref{Operation-Percentages} shows the relative frequency of selected HDFS operations in a workload generated by Hadoop applications, such as, \emph{Pig}, \emph{Hive}, \emph{HBase}, \emph{MapReduce}, \emph{Tez}, \emph{Spark}, and \emph{Giraph} at Spotify. \emph{List}, \emph{stat} and \emph{file read} operations alone account for $\approx 95\%$ of the operations in the HDFS cluster. These statistics are similar to the published workloads for Hadoop clusters at Yahoo~\cite{cristina14}, LinkedIn~\cite{indexFS}, and Facebook~\cite{BLOG_HDFS_CENTRAL_LOCK}. Figure~\ref{fig:pyramid}a shows the  relative cost of different database operations. We can see that the cost of a full table scan or an index scan, in which all database shards participate, is much higher than a partition pruned index scan in which only a single database shard participates. HopsFS metadata design and metadata partitioning enables implementations of common file system operations using only the low cost database operations, that is, primary key operations, batched primary key operations and partition pruned index scans. For example, the read and directory listing operations, are implemented using only (batched) primary key lookups and partition pruned index scans. Index scans and full table scans were avoided, where possible, as they touch all database shards and scale poorly. 

\subsection{Entity Relation Model}
In HopsFS, the file system metadata is stored in tables where a directory inode is represented by a single row in the \emph{Inode} table. File inodes, however, have more associated metadata, such as a set of blocks, block locations, and checksums that are stored in separate tables. 

Figure~\ref{fig:file-inode-related-metadata} shows the Entity Relational model depicting key entities in the HopsFS metadata model. Files and directories are represented by the \emph{Inode} entity that contains a reference to its parent inode (parent inode ID) in the file system hierarchy. We store path individual components, not full paths, in inode entries. Each file contains multiple blocks stored in the \emph{Block} entity. The location of each block replica is stored in the \emph{Replica} entity. During its life-cycle a block goes through various phases. Blocks may be under-replicated if a datanode fails and  such blocks are stored in the under-replicated blocks table (\emph{URB}). The replication manager, located on the leader namenode, sends commands to datanodes to create more replicas of under-replicated blocks. Blocks undergoing replication are stored in the pending replication blocks table (\emph{PRB}). Similarly, a replica of a block has various states during its life-cycle. When a replica gets corrupted, it is moved to the corrupted replicas (\emph{CR}) table. Whenever a client writes to a new block's replica, this replica is moved to the replica under construction (\emph{RUC}) table. If too many replicas of a block exist (for example, due to recovery of a datanode that contains blocks that were re-replicated), the extra copies are stored in the excess replicas (\emph{ER}) table and replicas that are scheduled for deletion are stored in the invalidation (\emph{Inv}) table. Note that the file inode related entities also contain the inode's foreign key (not shown in Figure~\ref{fig:file-inode-related-metadata}) that is also the partition key, enabling HopsFS to read the file inode related metadata using partition pruned index scans.

\subsection{Metadata Partitioning}
\label{sec:paritioning-metadata}

\begin{figure}[t]
 \centering
 \includegraphics[width=0.48\textwidth]{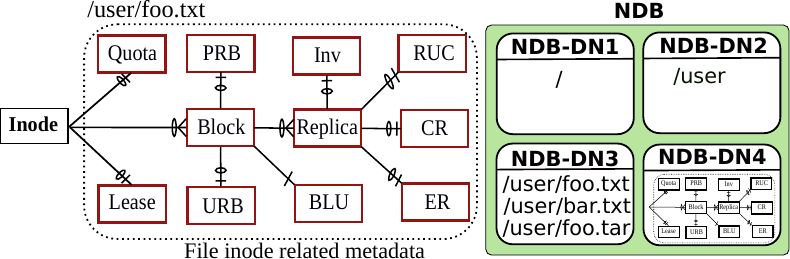}
 \caption{All the inodes in a directory are partitioned using a parent inode ID, therefore, all the immediate children of /user directory are stored on NDB-DN-3 for efficient directory listing, for example, ls /user. The file inode related metadata for /user/foo.txt is stored on NDB-DN-4 for efficient file reading operations, for example, cat /user/foo.txt.}
 \label{fig:file-inode-related-metadata}
\end{figure}

With the exception of hotspots (see the following subsection), HopsFS partitions \emph{inodes} by their \emph{parents' inode IDs}, resulting in inodes with the same parent inode being stored on the same database shard. This has the effect of uniformly partitioning the metadata among all database shards and it enables the efficient implementation of the directory listing operation. When listing files in a directory, we use a hinting mechanism to start the transaction on a transaction coordinator located on the database shard that holds the child inodes for that directory. We can then use a pruned index scan to retrieve the contents of the directory locally. File inode related  metadata, that is, blocks, replica mappings and checksums, is partitioned using the file's \emph{inode ID}. This results in metadata for a given file all being stored in a single database shard, again enabling efficient file operations, see Figure~\ref{fig:file-inode-related-metadata}.

\subsubsection{Hotspots}
\label{sec:hot-dirs}

A hotspot is an inode that receives a high proportion of file system operations. The maximum number of file system operations that can be performed on a 'hot' inode is limited by the throughput of the database shard that stores the inode. Currently, HopsFS does not have any built in mechanisms for identifying hotspots at run time.

All file system operations involve resolving the path components to check for user permissions and validity of the path. The  \emph{root} inode is shared among all file system valid paths. Naturally the \emph{root} inode is read by all file system path resolution operations. The database shard that stores the \emph{root} inode becomes a bottleneck as all file system operations will retrieve the \emph{root} inode from the same database shard. HopsFS caches the \emph{root} inode at all the namenodes. In HopsFS, the \emph{root} inode is immutable, that is, we do not allow operations, such as, renaming, deleting or changing the permissions of the \emph{root} inode. Making the \emph{root} inode immutable prevents any inconsistencies that could result from its caching. 

In HopsFS, all path resolution operations start from the second path component (that is, the top level directories). For the top-level directories, our partitioning scheme inadvertently introduced a hotspot -- all top-level directories and files are children of the root directory, and, therefore, resided on the same database shard. Operations on those inodes were handled by a single shard in the database. To overcome this bottleneck, HopsFS uses a configurable directory partitioning scheme where the immediate children of the top level directories are pseudo-randomly partitioned by hashing the names of the children. By default, HopsFS pseudo-randomly partitions only the first two levels of the file system hierarchy, that is, the root directory and its immediate descendants. However, depending on the file system workloads it can be configured to pseudo-randomly partition additional levels at the cost of slowing down \emph{move} and \emph{ls} operations at the top levels of the file system hierarchy.

\section{HopsFS Transactional Operations}
\label{sec-detailed-solution}

Transactional metadata operations in HopsFS belong to one of the two categories: \textbf{Inode} operations that operate on single file, directory or block (for example, create/read file, mkdir, and block state change operations), and \textbf{subtree} operations that operate on an unknown number of inodes, potentially millions, (for example, recursive \emph{delete}, \emph{move}, \emph{chmod}, and \emph{chown} on non-empty directories). 

This section describes how HopsFS efficiently encapsulates inode operations in transactions in NDB. The strongest transaction isolation level provided by NDB is \emph{read-committed}, which is not strong enough to provide at least as strong consistency semantics as HDFS which uses single global lock to serialize all HDFS operations. To this end, we use row-level locking to serialize conflicting inode operations. That is, the operations execute in parallel as long as they do not take conflicting locks on the same inodes. However, taking multiple locks in a transaction could lead to extensive deadlocking and transaction timeouts. The reasons are:\\
\noindent
\textbf{Cyclic Deadlocks:} In HDFS, not all inode operations follow the same order in locking the metadata which would lead to cyclic deadlocks in our case. To solve this problem, we have reimplemented all inode operations so that they acquire locks on the metadata in the same \textit{total order}, traversing the file system tree from the root down to leave nodes using left-ordered depth-first search.\\
\noindent
\textbf{Lock Upgrades:} In HDFS, many inode operations contain read operations followed by write operations on the same metadata. When translated into database operations within the same transaction, this results in deadlocking due to lock upgrades from read to exclusive locks. We have examined all locks acquired by the inode operations, and re-implemented them so that all data needed in a transaction is read only once at the start of the transaction (see Lock Phase,~\cref{sec:lock-phase}) at the strongest lock level that could be needed during the transaction, thus preventing lock upgrades. 



\subsection{Inode Hint Cache}
\label{sec:namenodes-cache}
Resolving paths and checking permissions is by far the most common operation in most HDFS workloads, see Table~\ref{Operation-Percentages}. In HDFS, the full path is recursively resolved into individual components. In HopsFS for a path of depth $N$, it would require $N$ roundtrips to the database to retrieve file path components, resulting in high latency for file system operations. 

Similar to AFS~\cite{AFS} and Sprite~\cite{Sprite}, we use \emph{hints}~\cite{LampsonHintsForComp} to speed up the path lookups. Hints are mechanisms to quickly retrieve file path components in parallel (batched operations). In our partitioning scheme, inodes have a composite primary key consisting of the parent inode's ID and the name of the inode (that is, file or directory name), with the parent inode's ID acting as the partition key. Each namenode caches only the primary keys of the inodes. Given a pathname and a hit for all path components directories, we can discover the primary keys for all the path components which are used to read the path components in parallel using a single database batch query containing only primary key lookups.

\subsubsection{Cache Consistency}
We use the inode hint cache entries to read the whole inodes in a single batch query at the start of a transaction for a file system operation. If a hint entry is invalid, a primary key read operation fails and path resolution falls back to recursive method for resolving file path components, followed by repairing the cache. Cache entries infrequently become stale as move operations, that update the primary key for an inode, are less than 2\% of operations in typical Hadoop workloads, see Table~\ref{Operation-Percentages}. Moreover, typical file access patterns follow a heavy-tailed distribution (in Yahoo 3\% of files account for 80\% of accesses~\cite{cristina14}), and using a $sticky$ policy  for HopsFS clients improves temporal locality and cache hit rates.
 
\subsection{Inode Operations}  
\label{sec:inode-and-block-ops}
HopsFS implements a pessimistic concurrency  model that supports parallel read and write operations on the namespace, serializing conflicting inode and subtree operations. We chose a pessimistic scheme as, in contrast to optimistic concurrency control, it delivers good performance for medium to high levels of resource utilization~\cite{Agrawal1987}, and many HDFS clusters, such as Spotify's, run at high load. Inode operations are encapsulated in a single transaction that consists of three distinct phases, which are, \textbf{lock}, \textbf{execute}, and \textbf{update} phases. 

\subsubsection{Lock Phase}
\label{sec:lock-phase}
In the lock phase, metadata is locked and read from the database with the strongest lock that will be required for the duration of the transaction. Locks are taken in the \emph{total order}, defined earlier. Inode operations are path-based and if they are not read-only operations, they only modify the last component(s) of the path, for example, \emph{rm /etc/conf} and \emph{chmod +x /bin/script}. Thus, only the last component(s) of the file paths are locked for file system operations. 

Figure~\ref{fig:tx-template} shows a transaction template for HopsFS inode operations. Using the inode hint cache the primary keys for the file path components are discovered, line~1. The transaction is started on the database shard that holds all or most of the desired data, line~2. A batched operation reads all the file path components up to the penultimate path component without locking (\emph{read-committed}) the metadata, line~3. For a path of depth \emph{N}, this removes \emph{N-1} round trips to the database. If the inode hints are invalid then the file path is recursively resolved and the inode hint cache is updated, line~4.

\floatstyle{plain}
\restylefloat{figure}
\floatstyle{boxed}
\restylefloat{figure}
\definecolor{opt}{rgb}{0.0, 0.51, 0.5}
\definecolor{tc}{rgb}{0.5, 0.0, 0.0} 
\definecolor{pc}{rgb}{0.47, 0.32, 0.66}
\definecolor{ec}{rgb}{0.25, 0.0, 1.0} 
\definecolor{err}{rgb}{1, 0, 0} 
\definecolor{rc}{rgb}{0.57, 0.0, 0.04} 

\begin{figure}[t]
\begin{Verbatim}[fontsize=\scriptsize,commandchars=\\\{\},codes={\catcode`$=3\catcode`^=7\catcode`_=8}]
 1. Get \textcolor{opt}{hints} from the inodes hint cache
 2. Set \textcolor{opt}{partition key hint} for the transaction
\textcolor{tc}{BEGIN TRANSACTION}
\textbf{\textcolor{pc}{LOCK PHASE:}}
 3. Using the inode \textcolor{opt}{hints}, batch read all \textcolor{ec}{inodes} 
    up to the penultimate \textcolor{ec}{inode} in the path
 4. If (cache miss || invalid path component) then
      \textcolor{err}{recursively} \textcolor{rc}{resolve} the path & update the cache
 5. Lock and \textcolor{rc}{read} the last \textcolor{ec}{inode}
 6. \textcolor{rc}{Read} \textcolor{ec}{Lease}, \textcolor{ec}{Quota}, \textcolor{ec}{Blocks}, \textcolor{ec}{Replica}, \textcolor{ec}{URB}, \textcolor{ec}{PRB}, \textcolor{ec}{RUC},
     \textcolor{ec}{CR}, \textcolor{ec}{ER}, \textcolor{ec}{Inv} using \textcolor{opt}{partition pruned index scans}
\textbf{\textcolor{pc}{EXECUTE PHASE:}}
 7. Process the data stored in the transaction cache
\textbf{\textcolor{pc}{UPDATE PHASE:}}
 8. Transfer the changes to database in batches
\textcolor{tc}{COMMIT/ABORT TRANSACTION}      
\end{Verbatim}
\vspace{-4mm}
\caption{Transaction template showing different optimization techniques, for example, setting a partition key hint to start a distribution aware transaction, inode hints to validate the file path components using a batch operation, and partition pruned index scans to read all file inode related metadata.}
\label{fig:tx-template}
\end{figure}

\floatstyle{plain}
\restylefloat{figure}

After the path is resolved, either a shared or an exclusive lock is taken on the last inode component in the path, line~5. Shared locks are taken for read-only inode operations, while exclusive locks are taken for inode operations that modify the namespace. Additionally, depending on the operation type and supplied operation parameters, inode related data, such as block, replica, and PRB, are read from the database in a predefined total order using partition pruned scans operations, line~6. 

HopsFS uses \emph{hierarchical locking}~\cite{gray76} for inode operations, that is, if data is arranged in tree like hierarchy and all data manipulation operations traverse the hierarchy from top to bottom, then taking a lock on the root of the tree/subtree implicitly locks the children of the tree/subtree. The entity relation diagram for file inode related data, see Figure~\ref{fig:file-inode-related-metadata}, shows that the entities are arranged in a tree with an inode entity at the root. That is, taking a lock on an inode implicitly locks the tree of file inode related data. As in all operations, inodes are read first, followed by its related metadata. For some operations, such as creating files/directories and listing operations, the parent directory is also locked to prevent \emph{phantom} and \emph{fuzzy} reads for file system operations.

\subsubsection{Per-Transaction Cache} All data that is read from the database is stored in a per-transaction cache (a snapshot) that withholds the propagation of the updated cache records to the database until the end of the transaction. The cache saves many round trips to the database as the metadata is often read and updated multiple times within the same transaction. Row-level locking of the metadata ensures the consistency of the cache, that is, no other transaction can update the metadata. Moreover, when the locks are released upon the completion of the transaction the cache is cleared. 


\subsubsection{Execute and Update Phases} The inode operation is performed by processing the metadata in the per-transaction cache. Updated and new metadata generated during the second phase is stored in the cache which is sent to the database in batches in the final \emph{update} phase, after which the transaction is either committed or rolled back. 

\section{Handling Large Operations}
\label{sec-subtree-ops}

Recursive operations on large directories, containing millions of inodes, are too large to fit in a single transaction, that is, locking millions of rows in a transaction is not supported in existing online~transaction processing systems. These operations include \emph{move}, \emph{delete}, \emph{change owner}, \emph{change permissions}, and \emph{set quota} operations. \emph{Move} operation changes the absolute paths of all the descendant inodes, while  \emph{delete} removes all the descendant inodes, and the \emph{set quota} operation affects how all the descendant inodes consume disk space or how many files/directories they can create. Similarly changing the permissions or owner of a directory may invalidate operations executing at the lower subtrees. 

\subsection{Subtree Operations Protocol}
Our solution is a protocol that implements subtree operations incrementally in batches of transactions. Instead of row level database locks, our \textit{subtree operations protocol} uses an application-level distributed locking mechanism to mark and isolate the subtrees.  We serialize subtree operations by ensuring that all ongoing inode and subtree operations in a subtree complete before a newly requested subtree operation is executed. We implement this serialization property by enforcing the following invariants: (1) no new operations access the subtree until the operation completes, (2) the subtree is quiesced before the subtree operation starts, (3) no orphaned inodes or inconsistencies arise if failures occur. 

Our subtree operations protocol provides the same consistency semantics as subtree operations in HDFS. For delete subtree operation HopsFS provides even stronger consistency semantics. Failed delete operations in HDFS can result in orphaned blocks that are eventually reclaimed by the block reporting subsystem (hours later). HopsFS improves the semantics of delete operation as failed operations does not cause any metadata inconsistencies, see~\cref{failed-sub-tree-ops}.
Subtree operations have the following phases.




\textbf{Phase 1:} In the first phase, an exclusive lock is acquired on the root of the subtree and a \emph{subtree lock} flag (which also contains the ID of the namenode that owns the lock) is set and persisted in the database. The flag is an indication that all the descendants of the subtree are  locked with exclusive (write) lock.

Before setting the lock it is essential that there are no other \emph{active subtree operations} at any lower level of the subtree. Setting the \emph{subtree lock} could fail active subtree operations executing on a subset of the subtree. We store all active subtree operations in a table and query it to ensure that no subtree operations are executing at lower levels of the subtree. 
In a typical workload, this table does not grow too large as subtree operations are usually only a tiny fraction of all file system operations.
It is important to note that during path resolution, inode and subtree operations that encounter an inode with a subtree lock turned on voluntarily abort the transaction and wait until the \emph{subtree lock} is removed. 

\begin{figure}[t]
 \centering
 \includegraphics[width=0.35\textwidth]{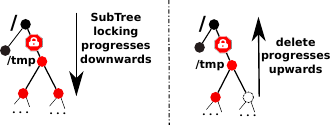}
 \caption{Execution of a delete subtree operation. Parallel transactions progress down (shown left) the subtree waiting for active operations to finish by taking and releasing write locks on all the descendant inodes. In the next phase (shown right), the delete operation is executed in batches using parallel transactions upwards from the leaf nodes.}

 \label{fig:subtree}
\end{figure}

\textbf{Phase 2:} To quiesce the subtree we wait for all ongoing inode operations to complete by taking and releasing database write locks on all inodes in the subtree in the same total order used to lock inodes. To do this efficiently, a pool of threads in parallel execute partition pruned index scans that write-lock child inodes. This is repeated down the subtree to the leaves, and, a tree data structure containing the inodes in the subtree is built in memory at the namenode, see Figure~\ref{fig:subtree}. The tree is later used by some subtree operations, such as, \emph{move} and \emph{delete} operations, to process the inodes. We reduce the overhead of reading all inodes in the subtree by using projections to only read the inode IDs.  If the subtree operations protocol fails to quiesce the subtree due to concurrent file system operations on the subtree, it is retried with exponential backoff.


\textbf{Phase 3:} In the last phase the file system operation is broken down into smaller operations that execute in parallel. For improved performance, large batches of inodes are manipulated in each transaction. 


\subsection{Handling Failed Subtree Operations}
\label{failed-sub-tree-ops} 
HopsFS takes lazy approach to cleanup subtree locks left by the failed namenodes~\cite{percolator}. Each namenode maintains a list of the active namenodes provided by the leader election service. If an operation encounters an inode with a subtree lock set and the namenode ID of the \emph{subtree lock}  belongs to a dead namenode then the \emph{subtree lock}  is cleared. However, it is important that when a namenode that is executing a subtree operation fails then it should not leave the subtree in an inconsistent state. The in-memory tree built during the second phase plays an important role in keeping the namespace consistent if the namenode fails. For example, in case of \emph{delete} operations the subtree is deleted incrementally in \emph{post-order} tree traversal manner using transactions. If half way through the operation the namenode fails then the inodes that were not deleted remain connected to the namespace tree. HopsFS clients will transparently resubmit the file system operation to another namenode to delete the remainder of the subtree. 

Other subtree operations (\emph{move}, \emph{set quota}, \emph{chmod} and \emph{chown}) do not cause any inconsistencies as the actual operation where the metadata is modified is done in the third phase using a single transaction that only updates the root inodes of the subtrees and the inner inodes are left intact. In the case of a failure, the namenode might fail to unset the \emph{subtree lock}, however, this is not a problem as other namenodes can easily remove the \emph{subtree lock} when they find out that the \emph{subtree lock} belongs to a dead namenode.

%

\subsection{Inode and Subtree Lock Compatibility}
Similar to the inode operation's locking mechanism (see~\cref{sec:lock-phase}), subtree operations also implement hierarchical locking, that is, setting a subtree flag on a directory implicitly locks the contents of the directory. Both inode and subtree locking mechanisms are compatible with each other, respecting both of their corresponding locks. That is, a subtree flag cannot be set on a directory locked by an inode operation and an inode operation voluntarily aborts the transaction when it encounters a directory with a subtree lock set.


\section{HopsFS Evaluation}
\label{hopsfs-performance}

As HopsFS addresses how to scale out the metadata layer of HDFS, all our experiments are designed to comparatively test the performance and scalability of the namenode(s) in HDFS and HopsFS in controlled conditions that approximate real-life file system load in big production clusters.

\subsection{Experimental Setup}

\textbf{Benchmark:} We have extended the benchmarking setup used to test the performance of Quantcast File System (QFS)~\cite{qfs}, which is an open source C++ implementation of Google File System. The benchmarking utility is a distributed application that spawns tens of thousands of HDFS/HopsFS file system clients, distributed across many machines, which concurrently execute file system (metadata) operations on the namenode(s). The benchmark utility can test the performance of both individual file system operations and file system workloads based on industrial workload traces. HopsFS and the benchmark utility are open source and the readers are encouraged to perform their own experiments to verify our findings~\cite{hammerBench, hopsfsGit}.


\textbf{HopsFS Setup:} All the experiments were run on premise using Dell PowerEdge R730xd servers(Intel(R) Xeon(R) CPU E5-2620 v3 @ 2.40GHz, 256 GB RAM, 4 TB 7200 RPM HDDs) connected using a single 10 GbE network adapter. Unless stated otherwise, NDB, version 7.5.3, was deployed on 12 nodes configured to run using 22 threads each and the data replication degree was 2.

\textbf{HDFS Setup:} In medium to large Hadoop clusters, 5 to 8 servers  are required to provide high availability for HDFS metadata service, see Figure~\ref{fig:hopsfs-arch} and section~\cref{sec:hopsfs-architecture}. The 5-server setup includes one active namenode, one standby namenode, at least three journal nodes collocated with at least three ZooKeeper nodes. In the 8-server setup, the ZooKeeper nodes are installed on separate servers to prevent multiple services from failing when a server fails. In our experiments Apache HDFS, version 2.7.2 was deployed on 5 servers. Based on Spotify's experience of running HDFS, we configured the HDFS namenodes with 240 client handler threads (\emph{dfs.namenode.handler.count}).

None of the file system clients were co-located with the namenodes or the database nodes. As we are only evaluating metadata performance, all the tests created files of zero length (similar to the NNThroughputBenchmark~\cite{shvachkoHdfsLimitations}). Testing with non-empty files requires an order of magnitude more HDFS/HopsFS datanodes, and provides no further insight.

\subsection{Industrial Workload Experiments}
\label{industrial-workloads}

We benchmarked HopsFS using workloads based on operational traces from Spotify that operates a Hadoop cluster consisting of $1600{\raisebox{1.8pt}{\scriptsize$+$}}$ nodes containing 60 petabytes of data. The namespace contains 13 million directories and 218 million files where each file on average contains 1.3 blocks. The Hadoop cluster at Spotify runs on average forty thousand jobs from different applications, such as, \emph{Pig}, \emph{Hive}, \emph{HBase}, \emph{MapReduce}, \emph{Tez}, \emph{Spark}, and \emph{Giraph} every day.  
The file system workload generated by these application is summarized in Table~\ref{Operation-Percentages}, which shows the relative frequency of HDFS operations.
At Spotify the average file path depth is 7 and average inode name length is 34 characters. On average each directory contains 16 files and 2 sub-directories. There are 289 million blocks stored on the datanodes. We use these statistics to generate file system workloads that approximate HDFS usage in production at Spotify.

\begin{figure}[t]
 \centering
\includegraphics[width=0.45\textwidth]{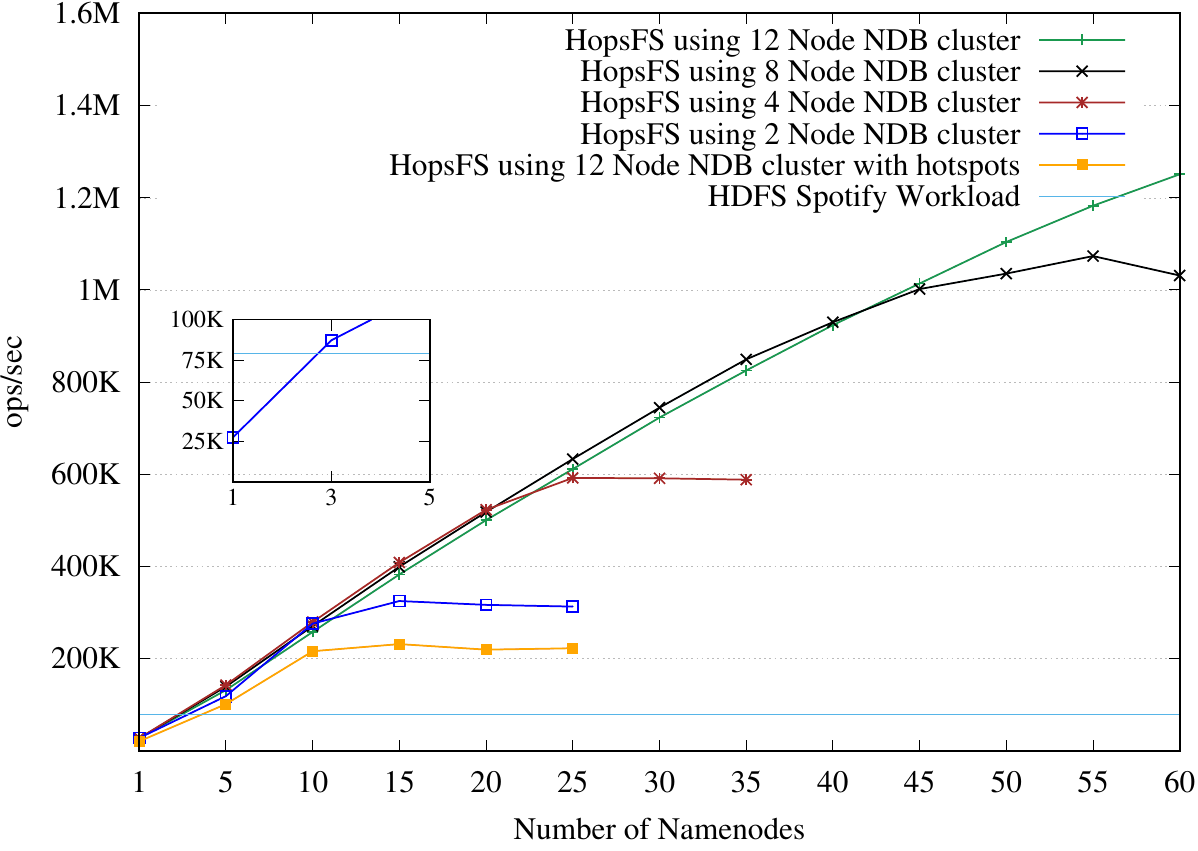}
 \caption{HopsFS and HDFS throughput for Spotify workload. }
 \label{fig:hops-throughput-interleaved}
\end{figure}

Figure~\ref{fig:hops-throughput-interleaved} shows that, for our industrial workload, using 60 namenodes and 12 NDB nodes, HopsFS can perform 1.25 million operations per second delivering \textbf{16 times} the throughput of HDFS. As discussed before in medium to large Hadoop clusters \textbf{5 to 8 servers} are required to provide high availability for HDFS. With equivalent hardware (2 NDB nodes and 3 namenodes), HopsFS delivers $\approx$10\% higher throughput than HDFS. HopsFS performance increases linearly as more namenodes nodes are added to the system. 

Table~\ref{write-intensive-workloads} shows the performance of HopsFS and HDFS for write intensive synthetic workloads. These synthetic workloads are derived from the previously described  workload, but here we increase the relative percentage of file create operations and reduce the percentage of file read operations. In this experiment, HopsFS is using 60 namenodes. As HopsFS only takes locks on inodes and subtrees, compared to HDFS' global lock, HopsFS outperforms HDFS by \textbf{37 times} for workloads where 20\% of the file system operations are file create operations. 

\begin{table}[hb]
\scriptsize
\centering
\begin{tabular}{|p{0.235\textwidth} p{0.05\textwidth} p{0.05\textwidth} p{0.06\textwidth}|}
\hline
\hline
\cellcolor{tableHeaderColor}\textbf{Workloads}    & \cellcolor{tableHeaderColor}\textbf{HopsFS ops/sec}     & \cellcolor{tableHeaderColor}\textbf{HDFS   ops/sec}   & \cellcolor{tableHeaderColor}\textbf{Scaling Factor} \\ 
\hline
\hline
\rowcolor{lightRowColor}\cellcolor{cellDarkHighlightColor}\textbf{Spotify Workload (2.7\% File Writes)}        & \cellcolor{cellDarkHighlightColor}\textbf{1.25 M}  &  \cellcolor{cellDarkHighlightColor}\textbf{78.9 K}    & \cellcolor{cellDarkHighlightColor}\textbf{16}      \\
\rowcolor{darkRowColor} Synthetic Workload (5.0\% File Writes)    & 1.19 M  &   53.6 K   &  \textbf{22} \\ 
\rowcolor{lightRowColor} Synthetic Workload (10\% File Writes)     & 1.04 M  &   35.2 K    & \textbf{30} \\
\rowcolor{darkRowColor} Synthetic Workload (20\% File Writes)     & 0.748 M &   19.9 K   & \textbf{37} \\ 
\hline    
\hline
\end{tabular}
\caption{HDFS and HopsFS Scalability for Different Workloads.}
\label{write-intensive-workloads}
\vspace{2mm}
\hrulefill
\end{table}

\subsubsection{Hotspots}
It is not uncommon for big data applications to create millions of files in a single directory~\cite{hdfs_workload, giga+}. As discussed in section~\ref{sec:hot-dirs} the performance of HopsFS is affected if the file system operations are not  uniformly distributed among all the database shards. In this experiment, all the file system operation paths share a common ancestor, that is, \emph{/shared-dir/...}. All the file system operations manipulate files and directories with common ancestor and the file system operations are generated using the workload described in the previous section~\ref{industrial-workloads}. The scalability of this workload is limited by the performance of the database shard that holds the \emph{/shared-dir}. Despite the fact that the current version of HopsFS does not yet provide a solution for scaling the performance of hotspots, the current solution outperforms HDFS by \textbf{3 times}, see Figure~\ref{fig:hops-throughput-interleaved}. We did not see any effect on the performance of HDFS in the presence of hotspots. 

\subsection{Metadata (Namespace) Scalability}
\label{metadata-scalability}

In HDFS, as the entire namespace metadata must fit on the heap of single JVM, the data structures are highly optimized to reduce the memory footprint~\cite{HADOOP-1687}. In HDFS, a file with two blocks that are replicated three ways requires \emph{\textbf{448 + L}} bytes of metadata\footnote{These size estimates are for HDFS version 2.0.4 from which HopsFS was forked. Newer version of HDFS require additional memory for new features such as snapshots and extended attributes.} where \emph{\textbf{L}} represents the filename length. If the file names are 10 characters long, then a 1 GB JVM heap can store 2.3 million files. In reality the JVM heap size has to be significantly larger to accommodate secondary metadata, thousands of concurrent RPC requests, block reports that can each be tens of megabytes in size, as well as other temporary objects.

\begin{table}[hb]
\scriptsize
\centering
\begin{tabular}{|p{0.07\textwidth} p{0.1\textwidth} p{0.1\textwidth}|}
\hline
\hline
\cellcolor{tableHeaderColor}    &     \multicolumn{2}{c|}{\cellcolor{tableHeaderColor}\textbf{Number of Files} }     \\ 
\cellcolor{tableHeaderColor}\textbf{Memory}    & \cellcolor{tableHeaderColor}\textbf{HDFS}     & \cellcolor{tableHeaderColor}\textbf{HopsFS}   \\ 
\hline
\hline
\rowcolor{lightRowColor} 1 GB			    & 2.3 million &        0.69 million  \\ 
\rowcolor{darkRowColor} 50 GB			    & 115 million &        34.5 million \\ 
\rowcolor{lightRowColor}100 GB			    & 230 million &        69 million \\ 
\rowcolor{darkRowColor} 200 GB			    & \cellcolor{cellDarkHighlightColor} \textbf{460 million} &        138 million \\ 
\rowcolor{lightRowColor}500 GB			    & Does Not Scale  &        346 million \\ 
\rowcolor{darkRowColor} 1 TB			    &  Does Not Scale  &        708 million \\ 
\rowcolor{lightRowColor} 24 TB			    &  Does Not Scale  &    \cellcolor{cellDarkHighlightColor}    \textbf{17 billion}\\ 
\hline    
\hline
\end{tabular}
\caption{HDFS and HopsFS Metadata Scalability.}
\label{metadata-scalability}
\vspace{2mm}
\hrulefill
\end{table}

Migrating the metadata to a database causes an expansion in the amount of memory required to accommodate indexes, primary/foreign keys and padding. In HopsFS the same file described above takes \textbf{1552 bytes} if the metadata is replicated twice. For a highly available deployment with an active and standby namenodes for HDFS, you will need twice the amount of memory, thus, HopsFS requires $\approx$ {\textbf{1.5} times more memory than HDFS to store metadata that is highly available. Table~\ref{metadata-scalability} shows the metadata scalability of HDFS and HopsFS.  


NDB supports up to 48 datanodes, which allows it to scale up to 24 TB of data in a cluster with 512 GB RAM on each NDB datanode. HopsFS can store up to 17 billion files using 24 TB of metadata, which is (\textbf{$\approx$37 times}) higher than HDFS. 

\subsection{FS Operations' Raw Throughput}

\begin{figure}[t]
\centering
\includegraphics[width=0.48\textwidth]{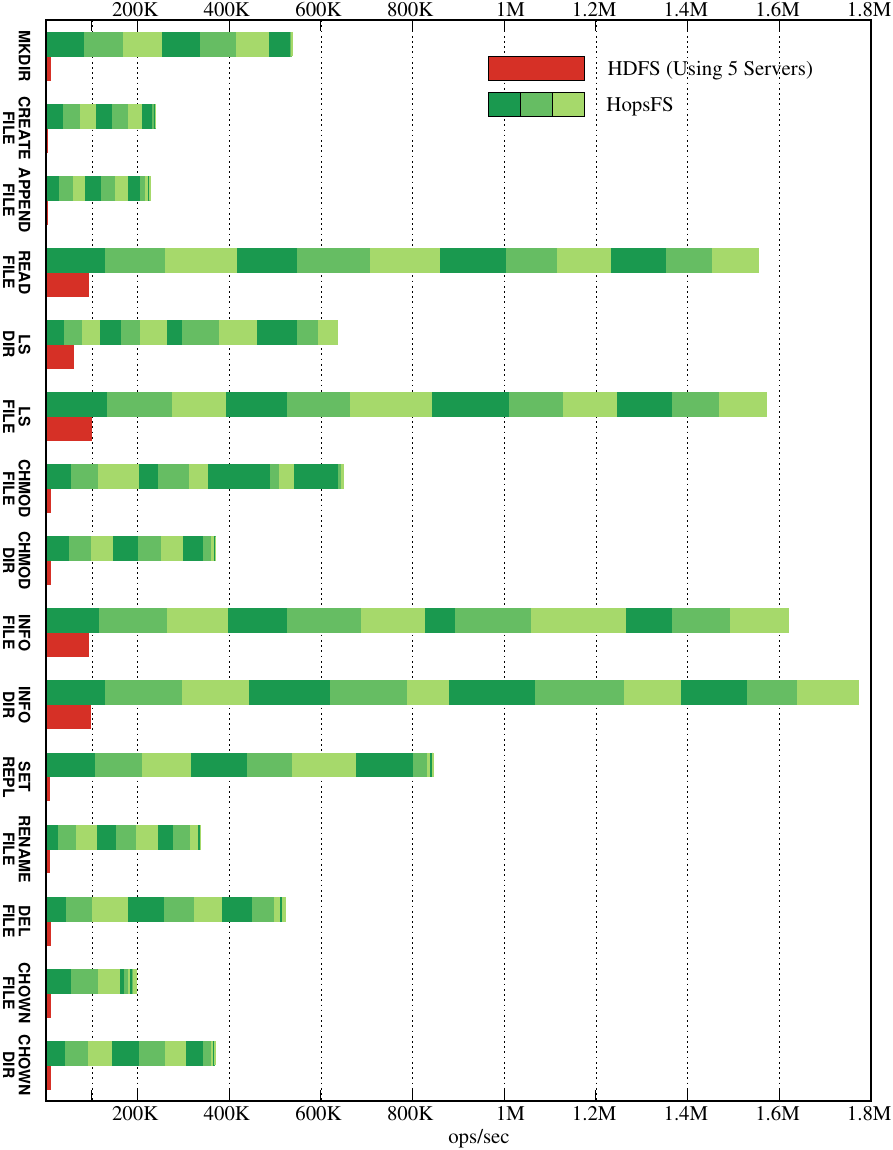}
\caption{HopsFS and HDFS throughput for different operations. For HopsFS each shaded box represents an increase in the throughput of the file system operation when five namenodes are added. For HDFS, the shaded box represents the maximum throughput achieved using the 5-server HDFS namenode setup. 
}
\label{fig:hops-throughput-raw}
\end{figure}

In this experiment, for each file system operation, the benchmark utility inundates the namenode(s) with the same file system operation. This test is particularly helpful in determining the maximum throughput and scalability of a particular file system operation. In real deployments, the namenode often receives a deluge of the same file system operation type, for example, a big job that reads large amounts of data will generate a huge number of requests to read files and list directories. 

Figure~\ref{fig:hops-throughput-raw} shows our results comparing the throughput for different file system operations. For each operation, HopsFS' results are displayed as a bar chart of stacked rectangles. Each rectangle represents an increase in the throughput when five new namenode are added. HopsFS outperforms HDFS for all file system operations and has significantly better performance than HDFS for the most common file system operations. 

\subsubsection{Subtree Operations}
In Table~\ref{table:sto-scalability}, we show the latency for \textit{move} and \textit{delete} subtree operations on a directory containing a varying number of files, ranging from one quarter to one million files. In this experiment, the tests were performed on HopsFS and HDFS clusters under 50\% load for the Spotify workload (50 \% of the maximum throughput observed in figure~\ref{fig:hops-throughput-interleaved}). 

In HopsFS, large amounts of data is read over the network and the operations are executed in many small transaction batches. The execution time of the move operation does not increase as rapidly because it does not update all the inner nodes or leaves of the subtree. HDFS outperforms HopsFS as all the data is readily available in the memory. However, due to the low frequency of such operations in typical industrial workloads (see Table~\ref{Operation-Percentages}), we think it is an acceptable trade-off for the higher performance of common file system operations in HopsFS.

\subsection{Operational Latency}
 
The latency for a single file system operation on an unloaded HDFS namenode will always be lower than in HopsFS, as all the metadata is readily available in main memory for the HDFS namenode, while it is remote for the namenodes in HopsFS. Figure~\ref{fig:hopsfs-latency-fn-of-clients} shows average file system operation latency observed by concurrent clients while running the Spotify workload. For such a workload, HopsFS has lower operation latency than HDFS because in HDFS file system operations that update the namespace block all other file system operations. Large HDFS deployments, may have tens of thousands of clients~\cite{shvachko_hdfs} and the end-to-end latency observed by the clients increases as the file system operations wait in RPC call queues at the namenode~\cite{fairHdfsRpc}. In contrast, HopsFS can handle more concurrent clients while keeping operation latencies low. 

\begin{table}[t]
\scriptsize
\centering
\begin{tabular}{|p{0.07\textwidth} p{0.07\textwidth} p{0.07\textwidth}p{0.07\textwidth} p{0.07\textwidth}|}
\hline
\hline
\cellcolor{tableHeaderColor}    &     \multicolumn{2}{c}{\cellcolor{tableHeaderColor}\textbf{mv} }    &     \multicolumn{2}{c|}{\cellcolor{tableHeaderColor}\textbf{rm -rf} }  \\ 
\cellcolor{tableHeaderColor}\textbf{Dir Size}    & \cellcolor{tableHeaderColor}\textbf{HDFS}     & \cellcolor{tableHeaderColor}\textbf{HopsFS}   & \cellcolor{tableHeaderColor}\textbf{HDFS} & \cellcolor{tableHeaderColor}\textbf{HopsFS}\\ 
\hline
\hline
\rowcolor{lightRowColor} 0.25 M			    & 197 ms & 1820 ms &  256 ms   & 5027 ms  \\ 
\rowcolor{darkRowColor} 0.50 M 			    & 242 ms & 3151 ms &  314 ms   & 8589 ms \\ 
\rowcolor{lightRowColor} 1.00 M			    & 357 ms & 5870 ms &  606 ms   & 15941 ms \\ 
\hline    
\hline
\end{tabular}
\caption{Performance of move and delete operations on large directories.}
\label{table:sto-scalability}
\vspace{2mm}
\hrulefill
\end{table}

\begin{figure}[thb]
 \centering
\includegraphics[width=0.48\textwidth]{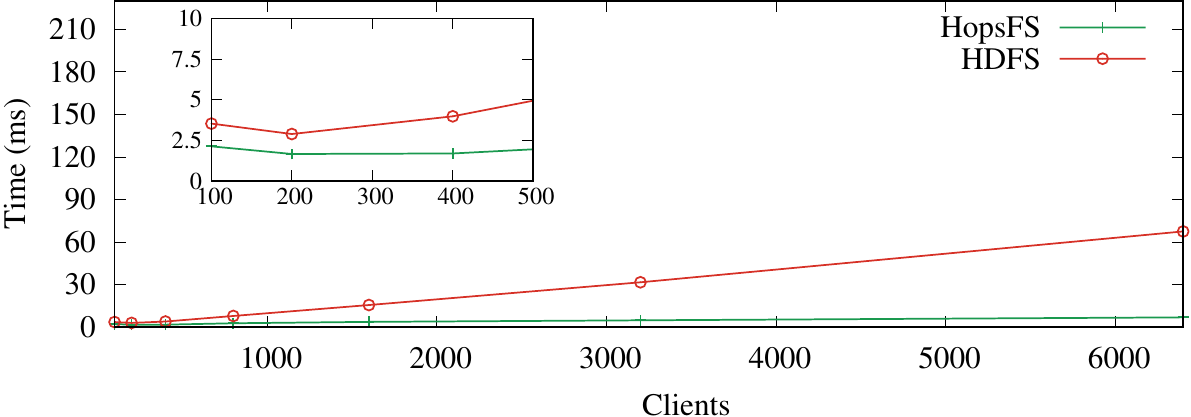}
 \caption{Average operation latency observed by HopsFS and HDFS for an increasing number of concurrent clients.}
 \label{fig:hopsfs-latency-fn-of-clients}
\end{figure}

Figure~\ref{fig:hopsfs-latency} shows 99th percentile latencies for different file system operations in a non-overloaded cluster. In this experiment, we ran HopsFS and HDFS under 50\% load for the Spotify workload (50 \% of the maximum throughput observed in Figure~\ref{fig:hops-throughput-interleaved}). 
In HopsFS, 99th-percentiles for common file system operations such as touch file, read file, ls dir and stat dir are 100.8 ms, 8.6 ms, 11.4  ms and 8.5 ms, respectively.  In a similar experiment for HDFS, running at 50\% load, the 99th-percentile latency for touch file, read file, ls dir and stat dir are 101.8, 1.5, 0.9, and  1.5 ms respectively.

\begin{figure}[t]
 \centering
\includegraphics[width=0.48\textwidth]{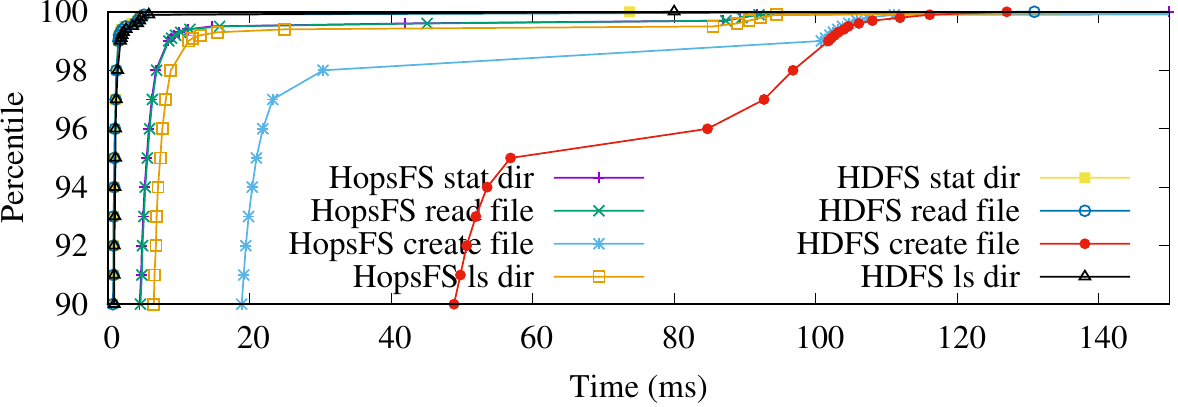}
 \caption{HopsFS and HDFS latency for common file system operations operating at 50\% load.}
 \label{fig:hopsfs-latency}
\end{figure}

\subsection{Failure Handling}

Now we discuss how the performance of the HDFS and HopsFS is affected when the namenodes, NDB datanodes, and journal nodes fail.

\subsubsection{Namenodes failure}
 
Figure~\ref{fig:hopsfs-failover} shows how the performance of the file system metadata service is affected when a namenode fails at 50\% of the load of the Spotify workload. The namenodes failures were simulated by killing and restarting all the file system processes on a namenode. For HDFS, the active namenode was periodically killed while for HopsFS, the namenodes were periodically killed in a round-robin manner. In the Figure~\ref{fig:hopsfs-failover}, vertical lines indicate namenode failures.
In HDFS, the standby namenode takes over when it detects that the active namenode has failed. In our experiments we have observed 8 to 10 seconds of downtime during failover in HDFS. During this time no file system metadata operation can be performed. Our failover tests were favorable to HDFS, as the amount of metadata stored by NNs in the experiment is minimal. At Spotify, with 140 gigabytes of metadata and 40 thousand jobs every day,  failover takes at least 5 minutes, and often up to 15 minutes. Although we are unsure of the reason why, it may be due to the additional checkpointing role played by the Standby namenode. Moreover, starting a namenode takes tens of minutes to build the in-memory representation of the name space from the on disk name space image and applying outstanding redo logs. 
In contrast, in HopsFS when a namenode fails clients transparently re-execute failed file system operations on one of the remaining namenodes in the system. In these experiments the number of file system clients were fixed and no new clients were added during the experiment. For HopsFS the throughput gradually drops as more and more namenodes are restarted. This is due to the fact that after a namenode fails the clients switch to remaining namenodes. In the experiments, HopsFS uses \emph{sticky} namenode selection policy and due to the fact that no new clients were started during the experiments the restarted namenodes do not receive as many file system operations requests as other namenodes.

\begin{figure}[ht]
 \centering
\includegraphics[width=0.48\textwidth]{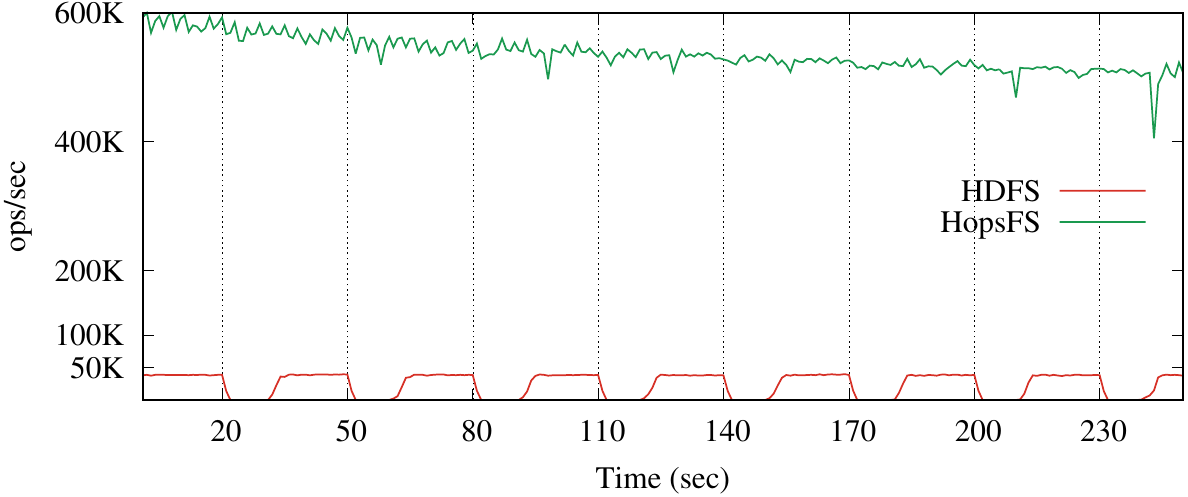}
 \caption{HopsFS and HDFS namenode failover. Vertical lines represent namenodes failures.}
 \label{fig:hopsfs-failover}
\end{figure}

\subsubsection{Failure of NDB Datanodes or Journal Nodes}

For a HDFS cluster with $N$ journal nodes, HDFS can tolerate failure of up to $\ceil{N/2}-1$ journal nodes. In our tests with a quorum of three journal nodes, HDFS can tolerate only one journal node failure. Increasing the size of the quorum to five enables HDFS to tolerate two journal nodes failure. 
We have tested HDFS with 3, 5, and 7 journal nodes, and the performance of the HDFS namenodes is not affected when the journal nodes fail provided that the quorum is not lost. When more journal nodes fail and the quorum is lost then the HDFS namenodes shutdown.

The number of NDB node failures that HopsFS can tolerate depends on the number of NDB datanodes and the replication degree. NDB is designed to provide 99.999\% availability~\cite{ndbAvailability}. With a default NDB replication degree of 2, a 4 node NDB cluster can tolerate up to 2 NDB datanodes failures and a 12 node NDB cluster can tolerate up to 6 NDB datanodes failure in disjoint replica groups. We have tested HopsFS on 2, 4, 8 and 12 node NDB clusters and the performance of HopsFS is not affected when a NDB datanode fails as long as there is at least one remaining NDB datanode alive in each node group. If all the NDB datanodes in a node group fail, then the HopsFS namenodes shutdown.

A common complaint against the two-phase commit protocol is that it is blocking and failure of a transaction coordinator or participant will cause the system to block. NDB internally implements a transaction coordinator failover protocol that hands over transactions on a failed coordinator to a different coordinator, (the default 1500 ms heartbeat timeout gives an upper bound of 6 seconds for 3 missed heartbeats). Transaction participant failures are identified by very low transaction inactive timeouts, (the default is 1200 ms also used in our experiments and in production). In the event of a transaction participant failure, failed transactions are automatically retried by the namenode and will be handled by the surviving datanodes in that replication group.

\subsection{Block Report Performance} 
In HDFS, each datanode periodically sends a block report to a namenode, containing IDs of all the blocks stored on the datanode. Block reports serve two purposes: (1) they help to rebuild the block location map when the namenode restarts since HDFS does not persist this information, (2) they serve as ground truth for available blocks in the system. We reimplemented the HDFS block reporting solution in HopsFS. Although the solution is fully functional it does not deliver as high throughput because a large amount of metadata is read over the network from the database by the namenodes to process a block report. 

In an experiment with the same setup, 150 datanodes simultaneously submitted block report containing $100,000$ blocks. With 30 namenodes, HopsFS manages to process 30 block reports per second while HDFS managed to process 60 block reports per second. However, full block-reports aren't needed as frequently in HopsFS as in HDFS, as we persist the block location mappings in the database. Even without further optimizations, with a 512 megabyte block size, and datanodes sending block reports every six hours, HopsFS can scale to handle block reporting in an exabyte cluster. 

\section{Related Work}
 
The InversionFS~\cite{inversionFs} and  Windows Future Storage (WinFS)~\cite{WinFS} were some of the first monolithic file systems that stored the metadata in a relational database.  Gunawi~\cite{gunawi2008sqck} showed that some file system operations, such as $fsck$, can be more efficient when implemented using a relational database. 

Recently, high performance distributed databases such as HBase \cite{hbaseBook,bigTable}, Cassandra~\cite{Cassandra}, CalvinDB~\cite{calvin} have enabled the development of new distributed metadata management architectures in file systems such as CalvinFS~\cite{CalvinFS}, CassandraFS~\cite{CassandraFS} and GiraffaFS~\cite{giraffa}. All of these file systems store denormalized metadata, that is, they store the full file path with each inode which affects the subtree operations. GiraffaFS only supports file move operation in the same directory. CalvinFS relies on CalvinDB to perform large transactions. CalvinDB runs large transactions in two phases. In the first phase the lock set is identified, and in the second phase all the locks are acquired and the operation is performed, provided that the lock set has not changed. However, CalvinFS did not experimentally show that this is a viable technique for performing operations on a directory with millions of files. Production-grade online transaction processing systems have an upper bound on the number of operations that can be included in a transaction, where the upper bound is much lower than tens of millions.

IndexFS~\cite{indexFS} and ShardFS~\cite{shardFS} are file systems optimized for metadata workloads with a large number of small files. IndexFS and ShardFS are middleware file systems, that is, they are built on existing distributed file systems such as HDFS~\cite{shvachko_hdfs}, Lustre~\cite{LustreV2}, PanFS~\cite{Panasas} and PVFS~\cite{pvfs2}. In IndexFS and ShardFS, the metadata servers  handle metadata as well as user data for small files stored in local LevelDB~\cite{levelDB} instances, and delegate the management of large files to an underlying distributed file system. For durability the LevelDB's SSTables are stored in the underlying distributed file system.  IndexFS caches inode information at clients, while ShardFS caches it at metadata servers. Atomic file system operations that involves both the underlying distributed file system and IndexFS/ShardFS metadata servers are not supported. For example, atomically deleting large files whose metadata is stored in the IndexFS/ShardFS metadata server and the file data is stored by the underlying distributed file system is not supported. IndexFS~\cite{indexFS} uses a caching mechanism to improve the performance of hot directories/files, while HopsFS' currently only load balances a user-configurable number of top-level directories. We are investigating more dynamic approaches for HopsFS. 

%

PVFS2~\cite{pvfs2}, OrangeFS~\cite{orangeFS}, Farsite~\cite{farsite2}, Lustre~\cite{LustreV2}, Vesta~\cite{Vesta}, InterMezzo~\cite{InterMezzo}, zFS~\cite{zfs}, and RAMA~\cite{RAMA} shard inodes among multiple metadata servers by either (1) random partitioning or (2) partition based hashed file identifiers or hashed full/partial file paths.  This partitioning scheme is typically combined with the caching of metadata at clients, which can cause cache invalidation storms for large subtree operations. Ceph dynamically partitions the file system tree, where hot-spot directories are hashed on multiple metadata servers~\cite{cephFS, cephMetadata}.

Finally, our architecture supports a pluggable NewSQL storage engine. MemSQL and SAP Hana are candidates, as they support high throughput cross-partition transactions, application defined partitioning, and partition pruned queries~\cite{memsql}. VoltDB is currently not a candidate as it serializes cross partition transactions~\cite{voldDBTimeout}. 


\section{External Metadata Implications}
\label{sec:implications-of-external-metadata}
Administrators often resort to writing their own tools to analyze the HDFS namespace. HopsFS enables online ad hoc analytics on the metadata. With a NDB backend, HopsFS metadata can be selectively and asynchronously replicated to either a backup cluster or a MySQL slave server, enabling complex analytics without affecting the performance of the active cluster. HopsFS metadata is also easy to export to external systems and it is easy to safely extend the metadata. That is, additional tables can be created that contain a foreign key to the associated inode, thus ensuring the integrity of the extended metadata. Using this approach, we have already added new features to HopsFS, including extended attributes for inodes and erasure coding. 
Moreover, following similar ideas to~\cite{HPStoreAllLazyBase}, we developed an eventually consistent replication protocol that replicates (extended) HopsFS metadata to Elasticsearch~\cite{elasticsearch} for free-text search. This enables us to search the entire namespace with sub-second latency. 
We believe that distributed metadata in a commodity database is a significant new enabling technology and it can become a reliable source of ground truth for metadata applications built on top of distributed file systems.

\section{Summary}
In this paper, we introduced HopsFS, that is, to the best of our knowledge, the first production-grade distributed hierarchical file system that stores its metadata in an external NewSQL database. HopsFS is an open-source, highly available file system that scales out in both capacity and throughput by adding new namenodes and database nodes. HopsFS can store 37 times more metadata than HDFS and for a workload from Spotify, HopsFS scales to handle 16 times the throughput of HDFS. HopsFS also has lower average latency for large number of concurrent clients, and no downtime during failover. 
Our architecture supports a pluggable database storage engine, and other NewSQL databases could be used. 
Finally, HopsFS makes metadata tinker friendly, opening it up for users and applications to extend and analyze in new and creative ways.


\section{Acknowledgements}
This work is funded by Swedish Foundation for Strategic Research project ``E2E-Clouds'', and  by EU FP7 project ``Scalable, Secure Storage and Analysis of Biobank Data'' under Grant Agreement no. 317871.

\balance
{
\footnotesize 
\bibliographystyle{abbrv}
\vspace{1mm}
\bibliography{hops-fs}

\begin{thebibliography}{10}

\bibitem{cristina14}
C.~L. Abad.
\newblock {\em {Big Data Storage Workload Characterization, Modeling and
  Synthetic Generation}}.
\newblock PhD thesis, University of Illinois at Urbana-Champaign, 2014.

\bibitem{Ursa_Minor}
M.~Abd-El-Malek, W.~V. Courtright, II, C.~Cranor, G.~R. Ganger, J.~Hendricks,
  A.~J. Klosterman, M.~Mesnier, M.~Prasad, B.~Salmon, R.~R. Sambasivan,
  S.~Sinnamohideen, J.~D. Strunk, E.~Thereska, M.~Wachs, and J.~J. Wylie.
\newblock {Ursa Minor: Versatile Cluster-based Storage}.
\newblock In {\em Proceedings of the 4th Conference on USENIX Conference on
  File and Storage Technologies - Volume 4}, FAST'05, pages 5--5, Berkeley, CA,
  USA, 2005. USENIX Association.

\bibitem{farsite1}
A.~Adya, W.~J. Bolosky, M.~Castro, G.~Cermak, R.~Chaiken, J.~R. Douceur,
  J.~Howell, J.~R. Lorch, M.~Theimer, and R.~P. Wattenhofer.
\newblock {Farsite: Federated, Available, and Reliable Storage for an
  Incompletely Trusted Environment}.
\newblock {\em SIGOPS Oper. Syst. Rev.}, 36(SI):1--14, Dec. 2002.

\bibitem{Agrawal1987}
R.~Agrawal, M.~J. Carey, and M.~Livny.
\newblock Concurrency control performance modeling: Alternatives and
  implications.
\newblock {\em ACM Trans. Database Syst.}, 12(4):609--654, Nov. 1987.

\bibitem{AlexandrovStratosphere}
A.~Alexandrov, R.~Bergmann, S.~Ewen, J.-C. Freytag, F.~Hueske, A.~Heise,
  O.~Kao, M.~Leich, U.~Leser, V.~Markl, F.~Naumann, M.~Peters,
  A.~Rheinl\"{a}nder, M.~J. Sax, S.~Schelter, M.~H\"{o}ger, K.~Tzoumas, and
  D.~Warneke.
\newblock {The Stratosphere Platform for Big Data Analytics}.
\newblock {\em The VLDB Journal}, 23(6):939--964, Dec. 2014.

\bibitem{megastore11}
J.~Baker, C.~Bond, J.~C. Corbett, J.~Furman, A.~Khorlin, J.~Larson, J.-M. Leon,
  Y.~Li, A.~Lloyd, and V.~Yushprakh.
\newblock Megastore: Providing scalable, highly available storage for
  interactive services.
\newblock In {\em Proceedings of the Conference on Innovative Data system
  Research (CIDR)}, pages 223--234, 2011.

\bibitem{txIosLvlCritique}
H.~Berenson, P.~Bernstein, J.~Gray, J.~Melton, E.~O'Neil, and P.~O'Neil.
\newblock A critique of ansi sql isolation levels.
\newblock {\em SIGMOD Rec.}, 24(2):1--10, May 1995.

\bibitem{CassandraFS}
{Cassandra File System Design}.
\newblock \url{http://www.datastax.com/dev/blog/cassandra-file-system-design}.
\newblock [Online; accessed 1-January-2016].

\bibitem{bigTable}
F.~Chang, J.~Dean, S.~Ghemawat, W.~C. Hsieh, D.~A. Wallach, M.~Burrows,
  T.~Chandra, A.~Fikes, and R.~E. Gruber.
\newblock {Bigtable: A Distributed Storage System for Structured Data}.
\newblock {\em {ACM} Trans. Comput. Syst.}, 26(2), 2008.

\bibitem{spanner}
J.~C. Corbett, J.~Dean, M.~Epstein, A.~Fikes, C.~Frost, J.~J. Furman,
  S.~Ghemawat, A.~Gubarev, C.~Heiser, P.~Hochschild, W.~Hsieh, S.~Kanthak,
  E.~Kogan, H.~Li, A.~Lloyd, S.~Melnik, D.~Mwaura, D.~Nagle, S.~Quinlan,
  R.~Rao, L.~Rolig, Y.~Saito, M.~Szymaniak, C.~Taylor, R.~Wang, and
  D.~Woodford.
\newblock {Spanner: Google's Globally-distributed Database}.
\newblock In {\em Proceedings of the 10th USENIX Conference on Operating
  Systems Design and Implementation}, OSDI'12, pages 251--264, Berkeley, CA,
  USA, 2012. USENIX Association.

\bibitem{Vesta}
P.~Corbett, D.~Feitelson, J.-P. Prost, and S.~Baylor.
\newblock {Parallel access to files in the Vesta filesystem}.
\newblock In {\em Supercomputing '93. Proceedings}, pages 472--481, Nov 1993.

\bibitem{EMC2_Hadoop}
E.~Corporation.
\newblock {HADOOP IN THE LIFE SCIENCES:An Introduction}.
\newblock
  \url{https://www.emc.com/collateral/software/white-papers/h10574-wp-isilon-hadoop-in-lifesci.pdf},
  2012.
\newblock [Online; accessed 30-Aug-2015].

\bibitem{DeanMR}
{Dean, Jeffrey and Ghemawat, Sanjay}.
\newblock Mapreduce: Simplified data processing on large clusters.
\newblock {\em Commun. ACM}, 51(1):107--113, Jan. 2008.

\bibitem{farsite2}
J.~R. Douceur and J.~Howell.
\newblock {Distributed Directory Service in the Farsite File System}.
\newblock In {\em Proceedings of the 7th Symposium on Operating Systems Design
  and Implementation}, OSDI '06, pages 321--334, Berkeley, CA, USA, 2006.
  USENIX Association.

\bibitem{elasticsearch}
{Elasticsearch}.
\newblock \url{https://www.elastic.co/products/elasticsearch}.
\newblock [Online; accessed 1-January-2016].

\bibitem{hbaseBook}
L.~George.
\newblock {\em {HBase: The Definitive Guide}}.
\newblock Definitive Guide Series. O'Reilly Media, Incorporated, 2011.

\bibitem{GFS-2003}
S.~Ghemawat, H.~Gobioff, and S.-T. Leung.
\newblock {The Google File System}.
\newblock {\em SIGOPS Oper. Syst. Rev.}, 37(5):29--43, Oct. 2003.

\bibitem{giraffa}
{GiraffaFS}.
\newblock \url{https://github.com/giraffafs/giraffa}.
\newblock [Online; accessed 1-January-2016].

\bibitem{gray76}
J.~Gray, R.~Lorie, G.~Putzolu, and I.~Traiger.
\newblock {Granularity of Locks and Degrees of Consistency in a Shared
  Database}.
\newblock In {\em IFIP Working Conference on Modelling in Data Base Management
  Systems}, pages 365--394. IFIP, 1976.

\bibitem{gunawi2008sqck}
H.~S. Gunawi, A.~Rajimwale, A.~C. Arpaci-Dusseau, and R.~H. Arpaci-Dusseau.
\newblock {SQCK: A Declarative File System Checker}.
\newblock In {\em Proc. of OSDI'08}, pages 131--146. USENIX Association, 2008.

\bibitem{hammerBench}
{Hammer-Bench: Distributed Metadata Benchmark to HDFS}.
\newblock \url{https://github.com/smkniazi/hammer-bench}.
\newblock [Online; accessed 1-January-2016].

\bibitem{hdfsGCPausesJira}
{Hadoop JIRA: Add thread which detects JVM pauses.}
\newblock \url{https://issues.apache.org/jira/browse/HADOOP-9618}.
\newblock [Online; accessed 1-January-2016].

\bibitem{Herlihy90}
M.~P. Herlihy and J.~M. Wing.
\newblock Linearizability: A correctness condition for concurrent objects.
\newblock {\em ACM Trans. Program. Lang. Syst.}, 12(3):463--492, July 1990.

\bibitem{hopsfsGit}
{Hadoop Open Platform-as-a-Service (Hops) is a new distribution of Apache
  Hadoop with scalable, highly available, customizable metadata.}
\newblock \url{https://github.com/smkniazi/hammer-bench}.
\newblock [Online; accessed 1-January-2016].

\bibitem{zookeeper}
P.~Hunt, M.~Konar, F.~P. Junqueira, and B.~Reed.
\newblock {ZooKeeper: Wait-free Coordination for Internet-scale Systems}.
\newblock In {\em Proceedings of the 2010 USENIX Conference on USENIX Annual
  Technical Conference}, USENIXATC'10, pages 11--11, 2010.

\bibitem{xtreemfs}
F.~Hupfeld, T.~Cortes, B.~Kolbeck, J.~Stender, E.~Focht, M.~Hess, J.~Malo,
  J.~Marti, and E.~Cesario.
\newblock {The XtreemFS architecture—a case for object-based file systems in
  Grids}.
\newblock {\em Concurrency and computation: Practice and experience},
  20(17):2049--2060, 2008.

\bibitem{isardDryad}
M.~Isard, M.~Budiu, Y.~Yu, A.~Birrell, and D.~Fetterly.
\newblock Dryad: Distributed data-parallel programs from sequential building
  blocks.
\newblock In {\em Proceedings of the 2Nd ACM SIGOPS/EuroSys European Conference
  on Computer Systems 2007}, EuroSys '07, pages 59--72, New York, NY, USA,
  2007. ACM.

\bibitem{HPStoreAllLazyBase}
C.~Johnson, K.~Keeton, C.~B. Morrey, C.~A.~N. Soules, A.~Veitch, S.~Bacon,
  O.~Batuner, M.~Condotta, H.~Coutinho, P.~J. Doyle, R.~Eichelberger, H.~Kiehl,
  G.~Magalhaes, J.~McEvoy, P.~Nagarajan, P.~Osborne, J.~Souza, A.~Sparkes,
  M.~Spitzer, S.~Tandel, L.~Thomas, and S.~Zangaro.
\newblock From research to practice: Experiences engineering a production
  metadata database for a scale out file system.
\newblock In {\em Proceedings of the 12th USENIX Conference on File and Storage
  Technologies}, FAST'14, pages 191--198, Berkeley, CA, USA, 2014. USENIX
  Association.

\bibitem{Cassandra}
A.~Lakshman and P.~Malik.
\newblock {Cassandra: A Decentralized Structured Storage System}.
\newblock {\em SIGOPS Oper. Syst. Rev.}, 44(2):35--40, Apr. 2010.

\bibitem{LampsonHintsForComp}
B.~W. Lampson.
\newblock Hints for computer system design.
\newblock {\em SIGOPS Oper. Syst. Rev.}, 17(5):33--48, Oct. 1983.

\bibitem{pvfs2}
R.~Latham, N.~Miller, R.~B. Ross, and P.~H. Carns.
\newblock {A Next-Generation Parallel File System for Linux Clusters}.
\newblock {\em LinuxWorld Magazine}, January 2004.

\bibitem{levelDB}
{LevelDB}.
\newblock \url{http://leveldb.org/}.
\newblock [Online; accessed 1-January-2016].

\bibitem{LevyDfsServey}
E.~Levy and A.~Silberschatz.
\newblock Distributed file systems: Concepts and examples.
\newblock {\em ACM Computing Surveys}, 22:321--374, 1990.

\bibitem{memsql}
{MemSQL, The World's Fastest Database, In Memory Database, Column Store
  Database}.
\newblock \url{http://www.memsql.com/}.
\newblock [Online; accessed 30-June-2015].

\bibitem{RAMA}
E.~L. Miller and R.~Katz.
\newblock {RAMA: An Easy-To-Use, High-Performance Parallel File System}.
\newblock {\em Parallel Computing}, 23(4):419--446, July 1997.

\bibitem{AFS}
J.~H. Morris, M.~Satyanarayanan, M.~H. Conner, J.~H. Howard, D.~S. Rosenthal,
  and F.~D. Smith.
\newblock {Andrew: A Distributed Personal Computing Environment}.
\newblock {\em Commun. ACM}, 29(3):184--201, Mar. 1986.

\bibitem{ndbAvailability}
{MySQL Cluster: High Availability}.
\newblock \url{https://www.mysql.com/products/cluster/availability.html}.
\newblock [Online; accessed 23-May-2016].

\bibitem{mySQLCluster}
{MySQL Cluster CGE}.
\newblock \url{http://www.mysql.com/products/cluster/}.
\newblock [Online; accessed 30-June-2015].

\bibitem{inversionFs}
M.~A. Olson and M.~A.
\newblock {The Design and Implementation of the Inversion File System}, 1993.

\bibitem{Sprite}
J.~K. Ousterhout, A.~R. Cherenson, F.~Douglis, M.~N. Nelson, and B.~B. Welch.
\newblock {The sprite network operating system}.
\newblock {\em IEEE Computer}, 21:23--36, 1988.

\bibitem{qfs}
M.~Ovsiannikov, S.~Rus, D.~Reeves, P.~Sutter, S.~Rao, and J.~Kelly.
\newblock {The Quantcast File System}.
\newblock {\em Proc. VLDB Endow.}, 6(11):1092--1101, Aug. 2013.

\bibitem{newsql14sigmod}
F.~\"{O}zcan, N.~Tatbul, D.~J. Abadi, M.~Kornacker, C.~Mohan, K.~Ramasamy, and
  J.~Wiener.
\newblock {Are We Experiencing a Big Data Bubble?}
\newblock In {\em Proceedings of the 2014 ACM SIGMOD International Conference
  on Management of Data}, pages 1407--1408, 2014.

\bibitem{giga+}
S.~V. Patil, G.~A. Gibson, S.~Lang, and M.~Polte.
\newblock {GIGA+: Scalable Directories for Shared File Systems}.
\newblock In {\em Proceedings of the 2Nd International Workshop on Petascale
  Data Storage: Held in Conjunction with Supercomputing '07}, PDSW '07, pages
  26--29, New York, NY, USA, 2007. ACM.

\bibitem{NFS}
B.~Pawlowski, C.~Juszczak, P.~Staubach, C.~Smith, D.~Lebel, and D.~Hitz.
\newblock {NFS Version 3 - Design and Implementation}.
\newblock In {\em In Proceedings of the Summer USENIX Conference}, pages
  137--152, 1994.

\bibitem{percolator}
D.~Peng and F.~Dabek.
\newblock Large-scale incremental processing using distributed transactions and
  notifications.
\newblock In {\em Proceedings of the 9th USENIX Symposium on Operating Systems
  Design and Implementation}, 2010.

\bibitem{InterMezzo}
{Peter Braam Braam and Michael Callahan}.
\newblock {The InterMezzo File System}.
\newblock In {\em In Proceedings of the 3rd of the Perl Conference, O'Reilly
  Open Source Convention}, Monterey, CA, USA, 1999.

\bibitem{Exabyte_Cern}
A.~J. Peters and L.~Janyst.
\newblock {Exabyte Scale Storage at CERN}.
\newblock {\em Journal of Physics: Conference Series}, 331(5):052015, 2011.

\bibitem{polato2014comprehensive}
I.~Polato, R.~R{\'e}, A.~Goldman, and F.~Kon.
\newblock {A comprehensive view of Hadoop research -- A systematic literature
  review}.
\newblock {\em Journal of Network and Computer Applications}, 46:1--25, 2014.

\bibitem{LOCUS-GJ-Popek}
G.~Popek and B.~J. Walker.
\newblock {\em {The LOCUS distributed system architecture}}.
\newblock MIT Press, 1985.

\bibitem{GlobalFS}
K.~W. Preslan, A.~Barry, J.~Brassow, R.~Cattelan, A.~Manthei, E.~Nygaard, S.~V.
  Oort, D.~Teigland, M.~Tilstra, and et~al.
\newblock {Implementing Journaling in a Linux Shared Disk File System}, 2000.

\bibitem{hdfs_workload}
K.~Ren, Y.~Kwon, M.~Balazinska, and B.~Howe.
\newblock Hadoop's adolescence: an analysis of hadoop usage in scientific
  workloads.
\newblock {\em Proceedings of the VLDB Endowment}, 6(10):853--864, 2013.

\bibitem{indexFS}
K.~Ren, Q.~Zheng, S.~Patil, and G.~Gibson.
\newblock {IndexFS: Scaling File System Metadata Performance with Stateless
  Caching and Bulk Insertion}.
\newblock In {\em Proceedings of the International Conference for High
  Performance Computing, Networking, Storage and Analysis}, SC '14, pages
  237--248, Piscataway, NJ, USA, 2014. IEEE Press.

\bibitem{zfs}
O.~Rodeh and A.~Teperman.
\newblock {zFS - a scalable distributed file system using object disks}.
\newblock In {\em Mass Storage Systems and Technologies, 2003. (MSST 2003).
  Proceedings. 20th IEEE/11th NASA Goddard Conference on}, pages 207--218,
  April 2003.

\bibitem{Ronstroem2005}
M.~Ronstr{\"o}m and J.~Oreland.
\newblock {Recovery Principles of MySQL Cluster 5.1}.
\newblock In {\em Proc. of VLDB'05}, pages 1108--1115. VLDB Endowment, 2005.

\bibitem{fairHdfsRpc}
{RPC Congestion Control with FairCallQueue}.
\newblock \url{https://issues.apache.org/jira/browse/HADOOP-9640}.
\newblock [Online; accessed 1-January-2016].

\bibitem{salmanLE}
G.~B. Salman~Niazi, Mahmoud~Ismail and J.~Dowling.
\newblock {Leader Election using NewSQL Systems}.
\newblock In {\em Proc. of DAIS 2015}, pages 158 --172. Springer, 2015.

\bibitem{CODA}
M.~Satyanarayanan, J.~J. Kistler, P.~Kumar, M.~E. Okasaki, E.~H. Siegel, David,
  and C.~Steere.
\newblock {Coda: A Highly available File System for a Distributed Workstation
  Environment}.
\newblock {\em IEEE Transactions on Computers}, 39:447--459, 1990.

\bibitem{GPFS}
{Schmuck, Frank and Haskin, Roger}.
\newblock {GPFS: A Shared-Disk File System for Large Computing Clusters}.
\newblock In {\em Proceedings of the 1st USENIX Conference on File and Storage
  Technologies}, FAST '02, Berkeley, CA, USA, 2002. USENIX Association.

\bibitem{Seltzer2009}
M.~Seltzer and N.~Murphy.
\newblock {Hierarchical File Systems Are Dead}.
\newblock In {\em Proceedings of the 12th Conference on Hot Topics in Operating
  Systems}, HotOS'09, pages 1--1, Berkeley, CA, USA, 2009. USENIX Association.

\bibitem{HADOOP-1687}
K.~Shvachko.
\newblock {Name-node memory size estimates and optimization proposal}.
\newblock \url{https://issues.apache.org/jira/browse/HADOOP-1687}, August 2007.
\newblock [Online; accessed 11-Nov-2014].

\bibitem{shvachko_hdfs}
K.~Shvachko, H.~Kuang, S.~Radia, and R.~Chansler.
\newblock The hadoop distributed file system.
\newblock In {\em Proceedings of the 2010 IEEE 26th Symposium on Mass Storage
  Systems and Technologies (MSST)}, MSST '10, pages 1--10, Washington, DC, USA,
  2010. IEEE Computer Society.

\bibitem{shvachkoHdfsLimitations}
K.~V. Shvachko.
\newblock {HDFS Scalability: The Limits to Growth}.
\newblock {\em login: The Magazine of USENIX}, 35(2):6--16, Apr. 2010.

\bibitem{hopsLulea}
{HOPS, Software-As-A-Service from SICS’S new datacenter}.
\newblock
  \url{https://www.swedishict.se/hops-software-as-a-service-from-sicss-new-datacenter}.
\newblock [Online; accessed 23-May-2016].

\bibitem{mapr}
M.~Srivas, P.~Ravindra, U.~Saradhi, A.~Pande, C.~Sanapala, L.~Renu,
  S.~Kavacheri, A.~Hadke, and V.~Vellanki.
\newblock {Map-Reduce Ready Distributed File System}, 2011.
\newblock US Patent App. 13/162,439.

\bibitem{BLOG_HDFS_CENTRAL_LOCK}
{The Curse of the Singletons! The Vertical Scalability of Hadoop NameNode}.
\newblock
  \url{http://hadoopblog.blogspot.se/2010/04/curse-of-singletons-vertical.html}.
\newblock [Online; accessed 30-Aug-2015].

\bibitem{LustreV2}
{The Lustre Storage Architecture}.
\newblock
  \url{http://wiki.lustre.org/manual/LustreManual20_HTML/UnderstandingLustre.html}.
\newblock [Online; accessed 30-Aug-2015].

\bibitem{Frangipani}
C.~A. Thekkath, T.~Mann, and E.~K. Lee.
\newblock {Frangipani: A Scalable Distributed File System}.
\newblock In {\em Proceedings of the Sixteenth ACM Symposium on Operating
  Systems Principles}, SOSP '97, pages 224--237, New York, NY, USA, 1997. ACM.

\bibitem{CalvinFS}
A.~Thomson and D.~J. Abadi.
\newblock {CalvinFS: Consistent {WAN} Replication and Scalable Metadata
  Management for Distributed File Systems}.
\newblock In {\em 13th USENIX Conference on File and Storage Technologies (FAST
  15)}, pages 1--14, Santa Clara, CA, Feb. 2015. USENIX Association.

\bibitem{calvin}
A.~Thomson, T.~Diamond, S.-C. Weng, K.~Ren, P.~Shao, and D.~J. Abadi.
\newblock {Calvin: Fast Distributed Transactions for Partitioned Database
  Systems}.
\newblock In {\em Proceedings of the 2012 ACM SIGMOD International Conference
  on Management of Data}, SIGMOD '12, pages 1--12, New York, NY, USA, 2012.
  ACM.

\bibitem{voldDBTimeout}
{VoltDB Documentation}.
\newblock \url{http://docs.voltdb.com/ReleaseNotes/}.
\newblock [Online; accessed 30-June-2015].

\bibitem{cephFS}
S.~A. Weil, S.~A. Brandt, E.~L. Miller, D.~D.~E. Long, and C.~Maltzahn.
\newblock {Ceph: A Scalable, High-performance Distributed File System}.
\newblock In {\em Proceedings of the 7th Symposium on Operating Systems Design
  and Implementation}, OSDI '06, pages 307--320, Berkeley, CA, USA, 2006.
  USENIX Association.

\bibitem{cephMetadata}
S.~A. Weil, K.~T. Pollack, S.~A. Brandt, and E.~L. Miller.
\newblock {Dynamic Metadata Management for Petabyte-Scale File Systems}.
\newblock In {\em Proceedings of the 2004 ACM/IEEE Conference on
  Supercomputing}, SC '04, pages 4--, Washington, DC, USA, 2004. IEEE Computer
  Society.

\bibitem{Panasas}
{Welch, Brent and Unangst, Marc and Abbasi, Zainul and Gibson, Garth and
  Mueller, Brian and Small, Jason and Zelenka, Jim and Zhou, Bin}.
\newblock {Scalable Performance of the Panasas Parallel File System}.
\newblock In {\em Proceedings of the 6th USENIX Conference on File and Storage
  Technologies}, FAST'08, Berkeley, CA, USA, 2008. USENIX Association.

\bibitem{WinFS}
{WinFS: Windows Future Storage}.
\newblock \url{https://en.wikipedia.org/wiki/WinFS}.
\newblock [Online; accessed 30-June-2015].

\bibitem{shardFS}
L.~Xiao, K.~Ren, Q.~Zheng, and G.~A. Gibson.
\newblock {ShardFS vs. IndexFS: Replication vs. Caching Strategies for
  Distributed Metadata Management in Cloud Storage Systems}.
\newblock In {\em Proceedings of the Sixth ACM Symposium on Cloud Computing},
  SoCC '15, pages 236--249, New York, NY, USA, 2015. ACM.

\bibitem{orangeFS}
S.~Yang, W.~B. Ligon~III, and E.~C. Quarles.
\newblock {Scalable distributed directory implementation on orange file
  system}.
\newblock {\em Proc. IEEE Intl. Wrkshp. Storage Network Architecture and
  Parallel I/Os (SNAPI)}, 2011.

\bibitem{ZahariaSpark}
M.~Zaharia, M.~Chowdhury, M.~J. Franklin, S.~Shenker, and I.~Stoica.
\newblock Spark: Cluster computing with working sets.
\newblock In {\em Proceedings of the 2Nd USENIX Conference on Hot Topics in
  Cloud Computing}, HotCloud'10, pages 10--10, Berkeley, CA, USA, 2010. USENIX
  Association.

\bibitem{partitionPruning}
M.~Zait and B.~Dageville.
\newblock Method and mechanism for database partitioning, Aug.~16 2005.
\newblock US Patent 6,931,390.

\end{thebibliography}
}

\end{document}